\documentclass[preprint2]{emulateapj}

\usepackage{natbib,amsmath, amsfonts, amssymb}

\begin{document}

\title{SLUG -- Stochastically Lighting Up Galaxies I: Methods and Validating Tests
}

\author{
Robert L.  da Silva\altaffilmark{1,2},
Michele Fumagalli\altaffilmark{1}, 
and Mark Krumholz\altaffilmark{1}
}
\altaffiltext{1}{Department of Astronomy and Astrophysics, UCO/Lick Observatory, University of California, 1156 High Street, Santa Cruz, CA 95064}
\altaffiltext{2}{NSF Graduate Research Fellow}

\begin{abstract} 
The effects of stochasticity on the luminosities of stellar populations are an often neglected but crucial 
element
for understanding populations in the low mass or low star formation rate regime.
To address this issue, we present SLUG, a new code to ``Stochastically Light Up Galaxies". 
SLUG synthesizes stellar populations using a Monte Carlo technique that treats stochastic sampling properly
including the
%
effects of clustering, the stellar initial mass function, star formation history, stellar evolution, and 
cluster disruption. This code produces many useful outputs, including
 i) catalogs of star clusters and their properties, such as their stellar initial mass distributions and their photometric properties in a variety of filters, ii) two dimensional histograms of color-magnitude diagrams of every star in the simulation, iii) and the photometric properties of field stars and the integrated photometry of the entire simulated galaxy.
%
After presenting the SLUG algorithm in detail, we validate the 
code through comparisons with \textsc{SB99} in the well-sampled regime,
 and with observed photometry of Milky Way clusters. Finally, we demonstrate
the SLUG's capabilities by presenting outputs in the stochastic regime.
%
 SLUG is publicly distributed through the website http://sites.google.com/site/runslug/. 
\end{abstract}

\section{Introduction}\label{sec:intro}
Fundamental progress in understanding the properties of
galaxies, star clusters and  stellar populations 
comes from the comparisons between observed photometry  and  synthetic photometry derived from stellar evolution 
codes. It has become common practice to infer properties such as star formation rate (SFR), star formation history (SFH), age, metallicity,
redshift, and stellar mass from photometry.  Despite the limits of theoretical modeling of stellar populations
\citep[such as uncertainties with dust, stellar evolution, and the stellar initial mass function (IMF); see][]{conroy1, conroy2, conroy3} synthetic libraries have reached a degree of 
precision that allows accurate estimates of these parameters -- although sometimes with degeneracy -- 
in massive galaxies and clusters. 

However, observations reveal a higher complexity in lower mass systems where scaling relations
which apply to more massive systems cannot be trivially extrapolated 
\citep[e.g.,][]{lee2007, weisz}.  Moreover, in lower mass systems, the limited number of stars that 
are present  invalidates the basic assumption used by most of the currently available codes
for synthetic photometry (such as \textsc{starburst99} \citep[SB99;][]{starburst99}; PEGASE \citep{pegase}; and GALEV \citep{galev}): that the
IMF is fully sampled at all times. Violation of this assumption leads to
stochastic variations in photometric properties that these codes do not fully capture. 
%

For example in globular clusters, the simplest observed stellar populations, 
failure to account for sampling effects can lead to a dramatic overestimate of the 
 contributions of blue horizontal branch and AGB stars to the integrated light. As a result,
 correct estimates of globular cluster ages and metallicities based on their integrated light are possible 
 only if one correctly accounts for stochasticity
  \citep{colucci2011}.


Moreover, in weakly star forming regions, stochastic effects can mimic those of 
a varying IMF. Indeed, recent observations in the low SFR regime have led to
 serious consideration of a varying IMF \citep{pflamm2008, hoversten, meurer, lee2009}. 
 However a fully self-consistent model of stochasticity, allowing for a full range of parameters 
 such as differing degrees of stellar clustering, metallicities, stellar tracks,
  input IMFs and CMFs, and SFHs 
  has not been available to test the null hypothesis of a non-varying but stochastically sampled IMF. 

These considerations apply not only to the dwarf galaxies studied by \cite{lee2009} but also 
to the outer regions of galaxies
such as XUV disks \citep{boissier2007, thilker} and outlying \ion{H}{2} regions \citep{werk2008,gogarten} where
stochasticity becomes crucial in the interpretation of inferred SFRs and SFHs.

While the number of studies that use Monte Carlo approaches to address problems on scales of clusters and galaxies
is growing, 
  a general purpose tool to study photometry 
in clusters and galaxies has not previously been available. To fill this need,
we have created SLUG, a code to allow proper study of the stochastic star formation regime
 at a range of scales from individual star clusters to entire galaxies.
SLUG provides a variety tools for studying the stochastic regime,
such as the ability to create catalogs of clusters including their individual IMFs and photometric properties,
color-magnitude diagrams (CMDs) of entire galaxies where we keep track of the photometry of every star, as well
as integrated photometry of entire composite populations. 

This paper, the first of a series, 
focuses on the methods used in the code along with several tests to demonstrate that we are reliably reproducing
observations and other synthetic photometry predictions. We then demonstrate the use of this code in the stochastic regime.
 In a companion paper \citep{mikiletter}, 
we use SLUG to demonstrate that, once random sampling is included, a stochastic non-varying IMF can reproduce the observed
variation of the H$\alpha$/FUV ratio in dwarf galaxies without resorting to modifications of the IMF. 
In a the second paper of the series (da Silva et al in prep.) we will explore in 
detail the implications of stochastic star formation with clustering.
 Further work will apply this code to a variety of astrophysical questions, such as 
understanding SFR calibrations in the stochastic regime and further study of other claims of
a varying IMF.

The layout of the paper is as follows: $\S$\ref{sec:stoch} presents an introduction to stochasticity and its
effects on the luminosity of stellar populations;
$\S$\ref{sec:tech} gives a detailed description of the SLUG algorithm;  $\S$\ref{sec:tests} discusses various tests of the code; $\S$\ref{sec:inaction} 
shows a presentation of the code's outputs in the stochastic regime;
finally, $\S$\ref{sec:summary} summarizes the results.

\section{What is Stochasticity?}\label{sec:stoch}
Many astrophysical studies require creation of
synthetic photometry of galaxies and other collections
of stars in order to compare with observations. 
In this section we present a discussion of the various
effects of stochasticity and the regimes in which they are important.

\subsection{Coeval Stellar Populations}
The standard procedure for calculating the luminosity from a coeval population of stars used by
the most popular implementations (such as SB99) is as follows. 
To find
the luminosity per unit mass of a coeval population in some band $\beta$ at a time $t$ 
after formation ($\ell_{\beta, coeval}(t)$), one simply integrates
the luminosity per unit mass of each star in that band  as a function of mass and time ($\ell_\beta(m,t)$) weighted by
the distribution of stellar masses (i. e. the IMF) $\frac{dN}{d\ln m}$:
\begin{equation}
\ell_{\beta, coeval}(t)=\int_{m_{min}}^{m_{max}} \ell_\beta (m, t)\frac{dN}{d\ln m}dm.
\end{equation}
Note that here we use a normalization of the IMF such that $\int_{m_{min}}^{m_{max}} \frac{dN}{d\ln m}dm=1$.

By performing this integral, these models assume an infinitely well-sampled 
IMF. As a result the above formula is mass-independent, meaning that $\ell_{\beta,coeval}$
can be scaled according to the total amount of stellar mass in a population (i.e. the luminosity
of a mass $M$ of stars  is simply $M \ell_{\beta,coeval}$).
Thus a given amount of mass $M$ will have a 1-to-1 mapping
to a particular luminosity $L$. However for small stellar populations, the assumption
of continuous sampling breaks down and effects of stochasticity can become important. Specifically,
stochastic effects create a statistical dispersion of luminosities that result
from a given mass $M$ of stars based entirely on the probabilistic sampling
of the mass distribution of stars. This is because each realization
of a given mass $M$ is built up with a different distribution of
stellar masses which, due to the non-linear dependence of luminosity
on stellar mass, yields a different luminosity.
We call this type of stochastic process  {\it sampling} stochasticity.

Perhaps the most important manifestation of sampling stochasticity is undersampling of the
upper end of the IMF. Since the IMF is steeply declining
with increasing stellar mass, the expectation value of a low mass population
drawing a massive star is small. As a result, the IMF in a low mass
population with few stars can appear truncated and have less luminosity
than a fully-sampled assumption would have predicted. This is due to the 
very super linear dependence of luminosity on stellar mass.

 One can roughly estimate the mass below which this effect is insignificant by calculating the
expectation value of obtaining a star above a given mass. We do so following the formalism of \cite{elmegreen00},
who find that the total mass  ($M$) required to expect a single star above a mass $m$ is 
\begin{equation}
M\sim 3\times10^3 \left(\frac{m}{100 M_\odot}\right)^{1.35}.
\end{equation}
This statement is
clearly dependent on one's choice of IMF. \cite{elmegreen00} uses a 
Salpeter IMF with a a lower limit of 0.3 $M_\odot$ and no upper limit. If one imposes an upper limit to the stellar
mass function, this relation turns over and asymptotically approaches the limit. However, for order-of-magnitude purposes here, we neglect
such consideration.
This  result implies that in order to reasonably expect even a single 120 $M_\odot$ star\footnote{Due to limitations of stellar evolutionary tracks,
this is the highest stellar mass SLUG can model and is a reasonable
guess for the highly uncertain absolute stellar mass limit. While some \citep[e.g.,][]{figer2005} suggest a value of $\sim 120-150 M_\odot$
others \citep{crowther} suggest it may be as high as 300 $M_\odot$.}, one would 
need at least a total mass sampled  of approximately $10^4 M_\odot\equiv M_{trunc}$. 
Thus this IMF truncation effect of sampling stochasticity can be ignored
for \emph{coeval} populations with masses $\gg$ $M_{trunc}$. 
For more reference on the limits
 of stochastic sampling, we recommend \cite{cervino1} and \cite{cervino2}. For
specific considerations to H$\alpha$ luminosity (one of the features of a stellar population most sensitive to stochasticity), see \cite{cervino3}.

Another manner in which stochastic sampling can manifest in coeval populations is for stars
going through particularly short-lived and luminous phases of evolution after
they leave the main sequence (e.g., AGB and blue horizontal branch stars). 
Since these phases are short,
only a very narrow range of masses is undergoing one of them at any given time.
Thus the exact sampling within that mass range can have a large impact
on the number of stars within that phase. 
As a result, a non-infinite population of stars can have additional random
scatter in luminosity even if $M>M_{trunc}$. 
This effect is more important in populations with little ongoing star formation
relative to their stellar mass
(otherwise new stars dominate the photometric properties of the population),
at specific ages when these post-main sequence populations contribute significantly
to the luminosity of the population \citep{colucci2011}.

\begin{figure*}
\epsscale{0.75}
\plotone{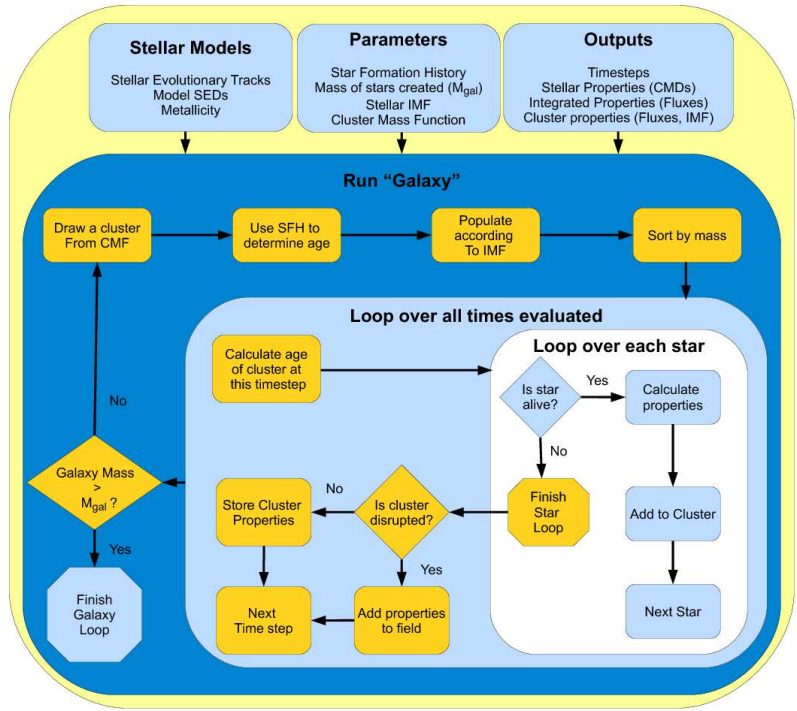}
\caption{A schematic flow-chart describing the algorithm of the
the SLUG code.  Note that for the case of unclustered star formation, the
cluster mass is drawn from the IMF and the population step is skipped as the single star is treated as part of a disrupted
cluster for the remainder of the code. Note this is updated from \cite{miki}.}
\label{fig:schematic}
\end{figure*}

\subsection{Composite Stellar Populations}\label{sec:comp}
In order to characterize a more complicated star formation history, 
SB99 and other such schemes  integrate over the coeval 
populations discussed above to find
the luminosity of all stars in a given band at a time $\tau$,
\begin{equation}
L_{\beta, total}(\tau)=\int_{-\infty}^\tau \mbox{SFR}(t) \ell_{\beta,coeval} ( \tau-t)dt.
\end{equation}

Such a treatment makes two key assumptions: (1) each of the summed coeval populations is 
large enough to ignore the effects of sampling stochasticity and (2) the SFR is 
 continuously sampled as well. These assumptions
can quickly break down for sufficiently low SFRs.

To illustrate this point, consider a galaxy forming stars at a constant rate. In order for the IMF  not to be truncated within some time
interval $dt$, there need to be at least $M_{trunc}$ worth of stars formed in that interval. For the SFR
to be considered reasonably well sampled, $dt$ must be much smaller than the evolutionary timescales 
of any of the stars, which are $\approx 10^6$ yr for the massive stars that generally dominate the light in an actively
star-forming system.
Thus these assumptions require
\begin{equation}\label{eqt:dt}
dt=\frac{M_{trunc}}{\mbox{SFR}} \ll 10^6 \mbox{yr}.
\end{equation}

\begin{deluxetable}{cl}
\tablewidth{0pc}
\tablecaption{Input Parameters}
\tablehead{\colhead{Parameter} & \colhead{Description}} 
\startdata
\cutinhead{Controlling the Physics}
IMF & stellar initial mass function; can choose \\
&Kroupa, Salpeter, Chabrier, IGIMF, or\\
& an arbitrary slope \vspace{1 mm}\\
CMF & cluster mass function, can change\\
&slope, minimum and maximum mass\vspace{1 mm}\\
Stellar Evolutionary  & library of models used for stellar evolution\\
Tracks &\vspace{1 mm}\\
Metallicity & metallicity of the stellar population\vspace{1 mm}\\
Stellar Atmosphere & which scheme and models are used for SEDs\vspace{1 mm}\\
Stellar Wind Model &  which wind model is used for SEDs\vspace{1 mm}\\
Fraction of stars in  & mass fraction of stars formed in clusters\\
clusters &\vspace{1 mm}\\
\cutinhead{Controlling the Simulation}
Maximum time & how long the simulation is run\vspace{1 mm}\\
SFH & can be arbitrary\vspace{1 mm}\\
Seed & random seed used for simulation\vspace{1 mm}\\
\cutinhead{Controlling Output}
Time step &  time between code outputs\vspace{1 mm}\\
Fluxes & choose which fluxes to output\vspace{1 mm}\\
Colors & which colors to use for CMDs\vspace{1 mm}\\
CMD output  & choice of number of bins and\\
parameters&range of color and luminosity for each CMD\vspace{1 mm}\\
Cluster output? & set to print output for each cluster\vspace{1 mm}\\
IMF output? & set to output IMF histograms for each cluster\vspace{1 mm}\\

\enddata
\label{table:inputs}
\end{deluxetable}

Thus these effects can only be ignored for SFRs consistently $\gg  10^{-2} M_\odot \mbox{yr}^{-1}\equiv\mbox{SFR}_{temp}$.
However, this \emph{temporal} stochasticity is amplified when one considers that stars are believed
to be formed in discrete collections known as clusters. As a result, 
the clumping in time of star formation in clusters can produce stochastic effects even in regions
with SFRs higher than SFR$_{temp}$. In this case the characteristic mass in
Equation \ref{eqt:dt} is replaced with a mass characteristic of the clusters being drawn
 (discussed further in da Silva in prep.; \citealt{mikiletter}).

The conditions required to treat a stellar population as continuous (as opposed to stochastically sampled)
break down in a variety of
astrophysical environments such as dwarf galaxies \citep[e.g.,][]{lee2009}, low star formation rate regions in the
outskirts of galaxies \citep[e.g.,][]{boissier07, miki08, bigiel10}, and low surface brightness galaxies \citep[e.g.,][]{boissier08}. 

\section{Technique}\label{sec:tech}
\subsection{Overview}

Here we present a brief overview of the code while we present each step in detail in the subsequent sections.

SLUG simulates star formation according to the scheme presented in Figure~\ref{fig:schematic}. We create
collections of star clusters obeying a user-defined cluster mass function (CMF) (which can include a given mass fraction of stars not formed in clusters),  SFH,  IMF, and choice of stellar evolutionary tracks, which we
call a ``galaxy". A description of the parameters that users can vary is provided in Table~\ref{table:inputs}.

These galaxies are built up  ($\S$\ref{sec:imf}) by first drawing the mass of an individual cluster from a CMF. This cluster's mass
is then filled up with stars according to an IMF. 
The age of the cluster is drawn from a distribution weighted by the given SFH. Each of the stars within the cluster is evolved
using a stellar evolutionary track combined with a model spectral energy distribution (SED) to determine a variety of integrated fluxes
corresponding to commonly used photometric filters ($\S$\ref{sec:seds}).

At a given set of time steps, these fluxes are summed over each star cluster. The clusters
are then disrupted according to the prescription of  \cite{fall2009}. Disrupted clusters have their fluxes added to a ``field" 
population while surviving clusters have their properties stored individually. 
The code repeats this process until a stellar mass equal to the integral of the provided SFH is created.

The code outputs a variety of  files that keep track of the properties of the stars, clusters, and total integrated stellar populations. Table~\ref{table:output} provides a short description of each available output file. All outputs are parsed and transformed into binary FITS tables.

\begin{figure}
\plotone{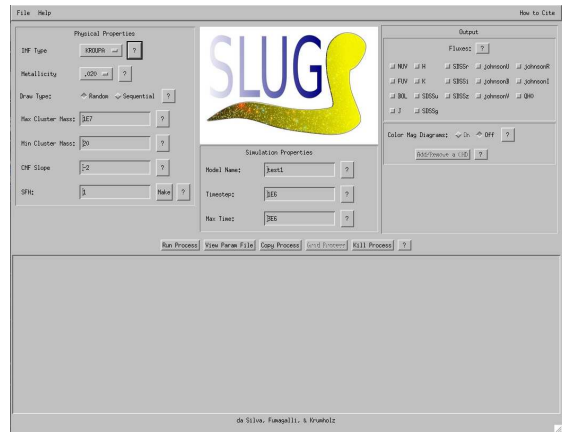}
\caption{ IDL GUI interface for running the code. The code may also be called via the UNIX or IDL command lines.}
\label{fig:gui}
\end{figure}

The code is open source and written in C++ with wrapping and parsing routines written in IDL.
This entire process can be controlled through an IDL graphical user interface (see Figure \ref{fig:gui}) or either of the UNIX or IDL command lines. The IDL routines are
wrapped in packages for use with the IDL virtual machine\footnote{which is available for free from http://www.ittvis.com/language/en-us/productsservices/idl/idlmodules/idlvirtualmachine.aspx} for those without IDL licenses.
For a full manual on how to use the code, visit the SLUG website at http://sites.google.com/site/runslug/.

\begin{deluxetable*}{cc}
\tablewidth{0pc}
\tablecaption{SLUG Output Files}
\tablehead{\colhead{Name} & \colhead{Description} } 
\startdata
Histogram & a 2d histogram of the user's choice of color-magnitude diagram(s)\\
& of every star in the ``galaxy" at each timestep\\
Cluster &  mass, fluxes, most massive star, number of stars, and \\
&age of each undisrupted cluster at each timestep\\
IMF & a histogram of the IMF of each cluster that appears in the Cluster file\\
Integral & the total flux of the entire ``galaxy" at each timestep\\
Miscellaneous &  the total stellar mass actually formed,\\
& as well as the actual SFH and CMF of the simulation\\
\enddata
\label{table:output}
\end{deluxetable*}

\subsection{Cluster Creation}\label{sec:imf}

\begin{figure}
\plotone{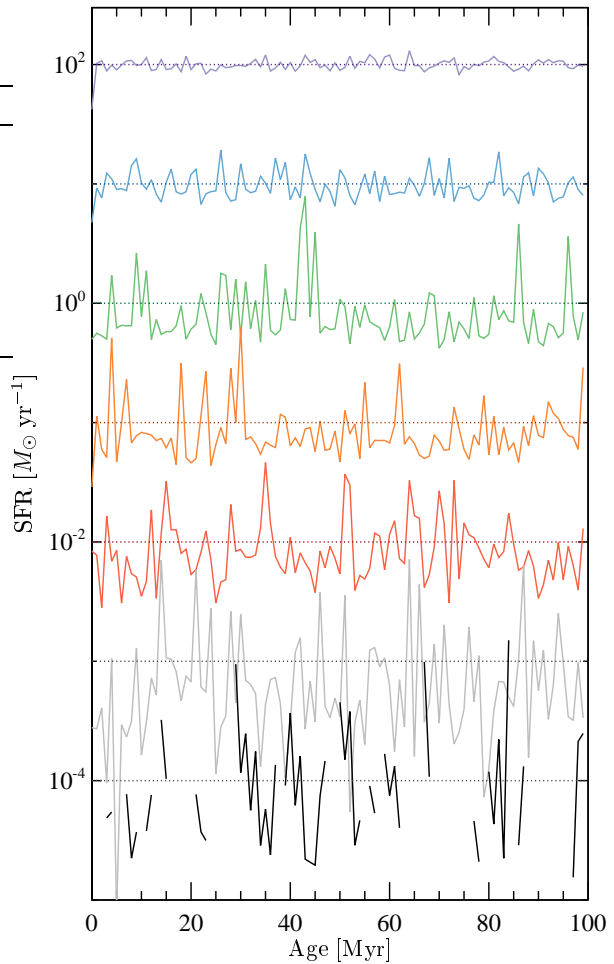}
\caption{Examples of star formation histories average over 1 Myr bins 
 for simulations with varying input constant SFRs of 0.0001--100 $M_\odot$ yr$^{-1}$.
The dotted lines show the input SFR. The average SFR of the simulation in each case is within 2, 0.2, and $<$0.02 percent
of the input for  $10^{-4}$,$10^{-3}$, and $>10^{-2}$ $M_\odot$ yr$^{-1}$ respectively.
SFRs of zero are masked. }
\label{fig:sfh}
\end{figure}
Most stars are thought to be born in star clusters \citep{ladalada} and
the distribution of star cluster masses 
appears to obey a power law distribution, where
observations \citep[e.g.,][]{zhang99,ladalada,fall09,chandar10} and theory \citep[e.g.,][]{fallkrummatz} suggest that the index ($\beta$) of the power law
$dN/dM\propto M^{-\beta}$ is approximately 2. SLUG allows for both clustered and unclustered star formation.
The user can choose what fraction of all stellar mass they wish to form in star clusters. If the code is forming
clusters, the CMF's power law slope as well as its upper and lower bounds can be varied. If
unclustered star formation is desired, the stars' masses are drawn individually from
an IMF and treated as a disrupted ``cluster"  of one star for the remainder of the code.

The initial masses of stars are drawn from an IMF. Choices of IMF\footnote{IMFs are truncated at
0.08$M_\odot$ due to lack of lower mass stellar tracks}  currently are \cite{chabrier},
\cite{Kroupa}, \cite{Salpeter}, a user-defined arbitrary power law, 
and the recently proposed IGIMF \citep{igimf0,pflamm2008}.
While the Chabrier, Kroupa, Salpeter, and power law IMFs are implemented
as a standard probability density function of stellar masses, the IGIMF has
additional features that require different treatment (see Appendix \ref{sec:igimf}).

Regardless of the choice of IMF, we draw stars until the total mass of the star cluster is built up. 
Since the random distribution of stars never exactly equals the mass of the cluster, a
question arises as to whether to keep the last star added. This last star
increases the mass of the cluster above the cluster mass drawn from the CMF. 
We determine whether or not to keep that star in the cluster based on whether keeping
the star in makes the total mass of stars closer to the mass drawn from the 
CMF than leaving it out\footnote{The effects of different sampling methods and 
their dependence on the CMF is studied in detail by \cite{haas2010}. Our method is
identical to their `stop-nearest' method.}.

Independent of its mass, the age of the cluster relative to the galaxy is assigned in a probabilistic manner weighted 
by the SFH (which can be arbitrary) such that the SFH is reproduced on average. 
This produces a scatter in the SFHs
for even a given ``constant" SFR.
 Thus SLUG's definition of a galaxy with a constant SFR is not a galaxy where the SFR is constant at every individual
time\footnote{A constant SFR cannot be instantaneously constant because stars form in discrete units of mass. For example, when a star is born, the instantaneous SFR is infinite, thus we must turn to a more probabilistic interpretation of the SFR.}, but rather a galaxy that produces an amount of stars over a time $dt$ equal to SFR$\times dt$ which are distributed in clusters
whose ages are drawn from a uniform distribution. This interpretation of what a SFR is and its
implications is discussed in more detail in da Silva et al. (in prep.). 

Clusters are born until the total mass
of stars formed is equal to the integral of the SFH. As with the problem of populating a
cluster with stars, a galaxy will never be filled to exactly its given mass with an integer number of
clusters. Therefore we apply the same condition for populating the galaxy as we do the clusters: we 
add until we exceed the mass and keep the final cluster only if the total galaxy mass is closer
to the desired value if we keep it. As a result the average SFR over the entire simulation of a particular galaxy can be 
higher or lower than the input value. This effect is small for most regimes, but very rare drawings of the CMF
at low SFRs can produce mild departures. We emphasize that this is not the effect of 
any error associated with the code but rather is the necessary result of our interpretation of what
a SFR means. 

We demonstrate the results of this procedure in Figure \ref{fig:sfh}. The figure shows that,
 while lower average SFRs tend to produce larger fractional scatter in the instantaneous SFR,
significant scatter remains until SFRs exceed 10 $M_\odot$ yr$^{-1}$.
This scatter is a direct result of the finite size of clusters. To clarify with an example, consider
that a  $10^7M_\odot$ cluster (when averaged
over the 1 Myr similarly to the curves shown in Figure \ref{fig:sfh}), will appear as a deviant peak for all but the highest
SFRs, where the contribution of that individual cluster is drowned out by enough other clusters.

We note that in this release of the code all stars in a cluster
are treated as having identically the same age. While observations suggest 
a scatter of a several Myr \citep{palla99, jeffries, hosokawa}, the mass dependence of this scatter is
unclear. 
Given the uncertainties, and that the intracluster age scatter is at most a few Myr, 
we chose to neglect this effect for now but plan on implementing it in the future.

\subsection{Stellar Tracks, SEDs, and Broad Band Photometry}\label{sec:seds}

Given the mass and age of each star, we need to determine its properties for a variety of observables. Our method
uses many of the same algorithms found in \textsc{SB99} \citep{starburst99, sb992} to create a 
set of tables from which SLUG interpolates. These tables are  constructed in advance so they need not be computed at run time.

Our first step is to determine the physical properties of each star.
To this end, we make use of a variety of stellar evolutionary models.
Modifying the \textsc{SB99} source code, we were able to obtain the full range of stellar tracks available to \textsc{SB99} (see Table~\ref{table:stellarprop}). In the future we plan to implement a wider range of
stellar tracks including those from \cite{eldridge} and the BaSTI library \citep{basti1, basti2}. We supplement the
Geneva tracks with the Padova+AGB tracks for stars in the mass range 0.15-0.8 $M_\odot$.
These models provide luminosities, gravities, chemical compositions, and effective temperatures at discrete intervals in the evolution of
a discrete number of stellar masses.  We then need to map these physical properties to stellar atmospheres in order to estimate the spectral energy distributions of the stars. Our code allows users to choose from one of five possible  \textsc{SB99} algorithms for modeling the atmospheres.
We implement all four prescriptions of stellar winds available in \textsc{SB99} (Maeder, empirical, theoretical, and Elson), which affect
the SEDs for Wolf-Rayet stars for some regimes and prescriptions. It is important to note that the SB99 algorithms
match SEDs to tracks with a nearest neighbor approach and not through interpolation. Therefore there can be 
some discreteness in the output SEDs. Future work will include removal of this effect.

With SEDs in hand, we can convolve with filters to determine the photometry of each point in our stellar tracks.
For this step we include the effects of nebular continuum (free-free, free-bound, and 2 photon processes) as implemented in SB99, but
 neglect nebular line emission for this first release of the code. (For a discussion of the importance of nebular continuum for the SEDs, see \citealt{nebcont}. Also see \citealt{nebcont3} and \citealt{nebcont2}.)
 The full list of available filters is presented in Table~\ref{table:filters}. We also integrate the SED to determine the bolometric luminosity as well as
to calculate Q(H$^{0}$), the number of hydrogen ionizing photons emitted per second.
One can convert Q(H$^{0}$) to H$\alpha$ luminosity with a simple conversion assuming
case B recombination (our notation follows \citealt{agn2}).
\begin{align}
L_{H\alpha}&=(1-f_{esc})(1-f_{dust})\mbox{Q(H}^0\mbox{)}\left(\frac{\alpha_{H\alpha}^{eff}}{\alpha_B}\right)h\nu_{H\alpha} \nonumber\\
    &\approx 1.37\times10^{-12}(1-f_{esc})(1-f_{dust})\mbox{Q(H}^0\mbox{)}\mbox{ ergs/s} 
\end{align}
where $f_{esc}$ is the escape fraction (thought to be between 0.05 \citep{boselli} and 0.4 \citep{hirashita}) and $f_{dust}$ represents the fraction of of ionizing photons  absorbed by dust grains 
\citep[e.g., see appendix of ][ who suggest a value of 0.37]{mckeewilliams97}.
 To better characterize the ionizing luminosity we also keep track of Q(He$^0$) and Q(He$^1$)
which represent the numbers of ionizing photons in the \ion{He}{1} and \ion{He}{2} continua respectively.

\begin{deluxetable*}{cc}
\tablewidth{0pc}
\tablecaption{Stellar Properties}
\tablehead{\colhead{Parameter} & \colhead{Allowed Values} } 
\startdata
{\bf Tracks} & Geneva STD\tablenotemark{a}, Geneva High\tablenotemark{a}, Padova STD\tablenotemark{b}, Padova AGB\tablenotemark{b}\\
{\bf Metallicity\tablenotemark{c} } & 0.0004-0.50 \\
{\bf SEDs} & Planck\tablenotemark{d}, Lejeune\tablenotemark{e}, Lejeune+Sch\tablenotemark{f}, Lejeune+SMI\tablenotemark{g}, Pau+SMI\tablenotemark{h}\\
{\bf Wind Models} &  Maeder\tablenotemark{i}, Empirical\tablenotemark{i}, Theoretical\tablenotemark{i}, Elson\tablenotemark{i} \\
\enddata
\label{table:stellarprop}
\tablenotetext{a}{\cite{meynet1994} and references therein}
\tablenotetext{b}{\cite{padova} and references therein}
\tablenotetext{c}{solar is 0.20}
\tablenotetext{d}{simple blackbody SED}
\tablenotetext{e}{\cite{lejeune1,lejeune2}}
\tablenotetext{f}{same as e, but for stars with strong winds use \cite{schmutz}}
\tablenotetext{g}{same as e, but for stars with strong winds use \cite{hillier}}
\tablenotetext{h}{same as g, but use \cite{pauldrach} for O stars}
\tablenotetext{i}{\cite{sb99winds}}
\end{deluxetable*}

The above steps allow us to create a discrete two-dimensional table for each
flux band where one axis represents stellar mass, the other represents
time, and the value of the table is the logarithm of
 the flux in that band at the appropriate mass and time. Our tables are created
 through use of the isochrone synthesis method such that
 our results are stable against the numerical issues that arise from a fixed mass approach \citep{isochrone}.

\begin{deluxetable}{cc}
\tablewidth{0pc}
\tablecaption{Broad Band Filters}
\tablehead{\colhead{Filter\qquad\qquad \qquad} & \colhead{Reference}} 
\startdata
NUV \qquad \qquad\qquad& 1 \\
FUV \qquad \qquad\qquad& 1 \\
$u$ \qquad\qquad \qquad& 2 \\
$g$\qquad \qquad\qquad & 2 \\
$r$ \qquad\qquad\qquad& 2\\
$i$\qquad \qquad\qquad & 2 \\
$z$ \qquad \qquad\qquad& 2 \\
J\qquad \qquad\qquad & 3 \\
H\qquad \qquad\qquad & 3 \\
K\qquad \qquad\qquad & 3 \\
U\qquad \qquad\qquad& 4\\
B \qquad \qquad\qquad& 4 \\
V\qquad \qquad\qquad& 4\\
R \qquad \qquad\qquad& 4 \\
I \qquad \qquad\qquad& 4 \\
Q(H$^0$)\qquad \qquad\qquad & 5 \\
Q(He$^0$) \qquad \qquad\qquad& 5 \\
Q(He$^1$) \qquad \qquad\qquad& 5 \\
$L_{\rm Bol}$ \qquad \qquad\qquad&6\\
\enddata
\tablenotetext{1}{\cite{galexfilt}}
\tablenotetext{2}{\cite{sdssfilt}}
\tablenotetext{3}{\cite{2mass}}
\tablenotetext{4}{\cite{forsfilt}}
\tablenotetext{5}{Obtained by integrating SED blueward of 912, 504, and 208 \AA\ for  Q(H$^0$), 
Q(He$^0$),
Q(He$^1$) respectively. }
\tablenotetext{6}{Given by stellar evolutionary tracks.}
\label{table:filters}
\end{deluxetable}

\begin{figure*}
\plotone{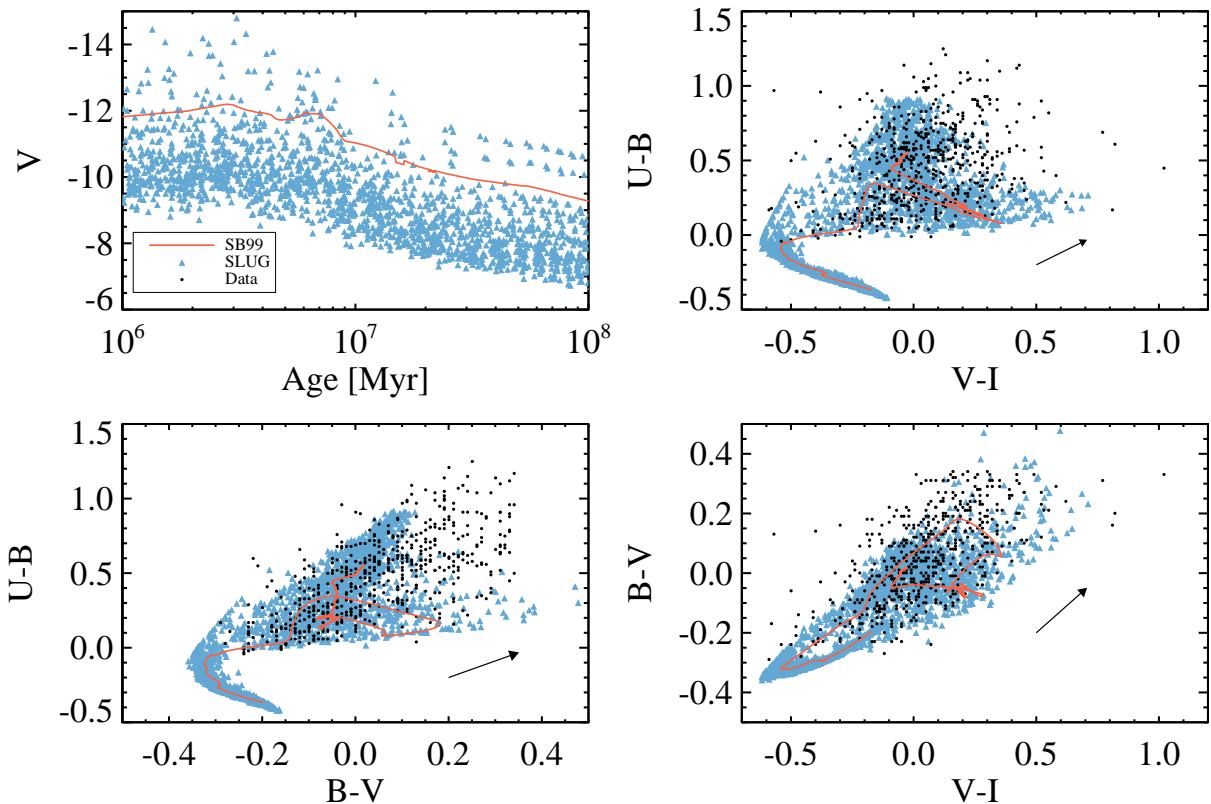}
\caption{Comparison of observed young star clusters from \cite{larsen} (black points) to SLUG models  of clusters $>10^4 M_\odot$ (blue triangles). 
The orange curve shows the
trajectory of a SB99 $10^5M_\odot$ cluster. Data are omitted from upper left panel as the ages are not present in the \cite{larsen} catalog. 
Arrows denote the 
extinction vector for $A_V=0.5$ mag  \citep[created following appendix B of][]{schlegel}.}
\label{fig:clusphot}
\end{figure*}

\subsection{Evaluating the Stellar Properties}

To determine the properties of a given star  of any mass at any given time, we first determine
if the star is still alive. This is done by an interpolation in time to find the minimum mass
of a dead star ($m_{death}$) at a given time according to our stellar evolution models (where we call
a star ``dead" if it no longer has entries in our stellar tracks). 
If the star is less massive than $m_{death}$, we
 interpolate our model tables to determine the flux in a given filter within 0.01 dex .

For computational speed, there are a variety of approximations and restrictions we are forced to implement.
The current scheme only allows ages up to 1 Gyr for the stellar tracks (to be expanded in later releases of the code).
We do not evolve stars less massive than 0.9 $M_\odot$ (a number which can be changed 
by the user). These stars do not evolve past the main sequence for the current maximum age of the code of 1 Gyr, so
these stars are treated as having their zero-age main sequence (ZAMS) properties.
 Due to limitations of the stellar tracks, we treat the photometric properties of all stars less massive than 0.156 $M_\odot$ identically
to those of 0.156 $M_\odot$ stars. For many purposes, more massive stars dominate the light
in the bands such that this approximation is reasonable. The tracks also impose a $120 M_\odot$ upper mass limit on stars.

Currently, we neglect the effects of binary stellar evolution \citep[see][]{eldridge}, which may have an impact
on the derived results by producing a bluer population with a reduced number of red supergiants and increased
age range of Wolf-Rayet stars.

\subsection{Cluster Disruption}
If the user chooses to form stars in star clusters, 
we randomly disrupt our clusters in a mass independent way such that $dN/d\tau\propto \tau^{-1}$ \citep[following][]{fall2009}. We
start cluster disruption 1 Myr after the cluster forms. This results in 90\% of star clusters
being disrupted  for each factor of 10 in age after 1 Myr. Stars in disrupted clusters still have their photometry calculated for
the integrated properties of the galaxy and are kept track in a set of ``field" variables and outputs.

\section{Validating Tests}\label{sec:tests}

\begin{deluxetable}{cc}
\tablewidth{0pc}
\tablecaption{Fiducial Inputs}
\tablehead{\colhead{Parameter} & \colhead{Fiducial Value}} 
\startdata
Time step & $10^6$ yr\\
Maximum time & $10^9$ yr\\
IMF & 1-120$M_\odot$; slope=-2.35\\
CMF & $20-10^7 M_\odot$; slope=-2\\
Stellar Evolutionary Tracks & Padova+AGB\\
Metallicity & Solar; $Z=0.20$\\
Stellar Atmosphere & Lej+Smi\\
Stellar Wind Model & Maeder\\
Fraction of stars in clusters & 100\%\\
\enddata
\label{table:fiducial}
\end{deluxetable}

In this section we present a variety of tests to validate the outputs of SLUG. For these tests we make use of a 
set of fiducial parameters presented in Table~\ref{table:fiducial} unless otherwise noted\footnote{While the preferred SEDs for \textsc{SB99} are the Pau+Smi atmospheres, we find that
the Pauldrach models are far too discrete. Therefore while we provide the Pau+Smi atmospheres, we recommend the Lej+Smi.}\footnote{Since we aim to test SLUG rather than to perform a study of the effects that the multiple parameters have on the luminosity distributions , we choose widely adopted vlaues.}.
To emphasize that SLUG can be applied at different scales, we arrange
these tests in order of scale starting with individual clusters and then considering integrated properties of entire galaxies in the well-sampled
regime.
\subsection{Photometry of Clusters}\label{sec:clusphot}

To demonstrate that SLUG reproduces properties of observed clusters, we turn to the
catalog of young star clusters compiled in \cite{larsen}. To reproduce the clusters we
modify our fiducial IMF to extend down to 0.08 $M_\odot$ and run a SLUG model
with a SFR of 1$M_\odot$ yr$^{-1}$ for 500 Myr, evaluated every 10 Myr.  
Note that the the SFR does not directly affect the CMF or the properties of the clusters, only the number of clusters
in existence at a given time. We show the results of this exercise in Figure~\ref{fig:clusphot} 
where we find
remarkable agreement between the models and the data. As is clear from the figure,
we are able to reproduce both the location and spread of most of the observed data. Clusters that
fall outside of the locus of SLUG models fall can easily be reproduced when one accounts for a 
modest amount of reddening (see reddening vector).

\begin{figure*}
\epsscale{0.9}
\plotone{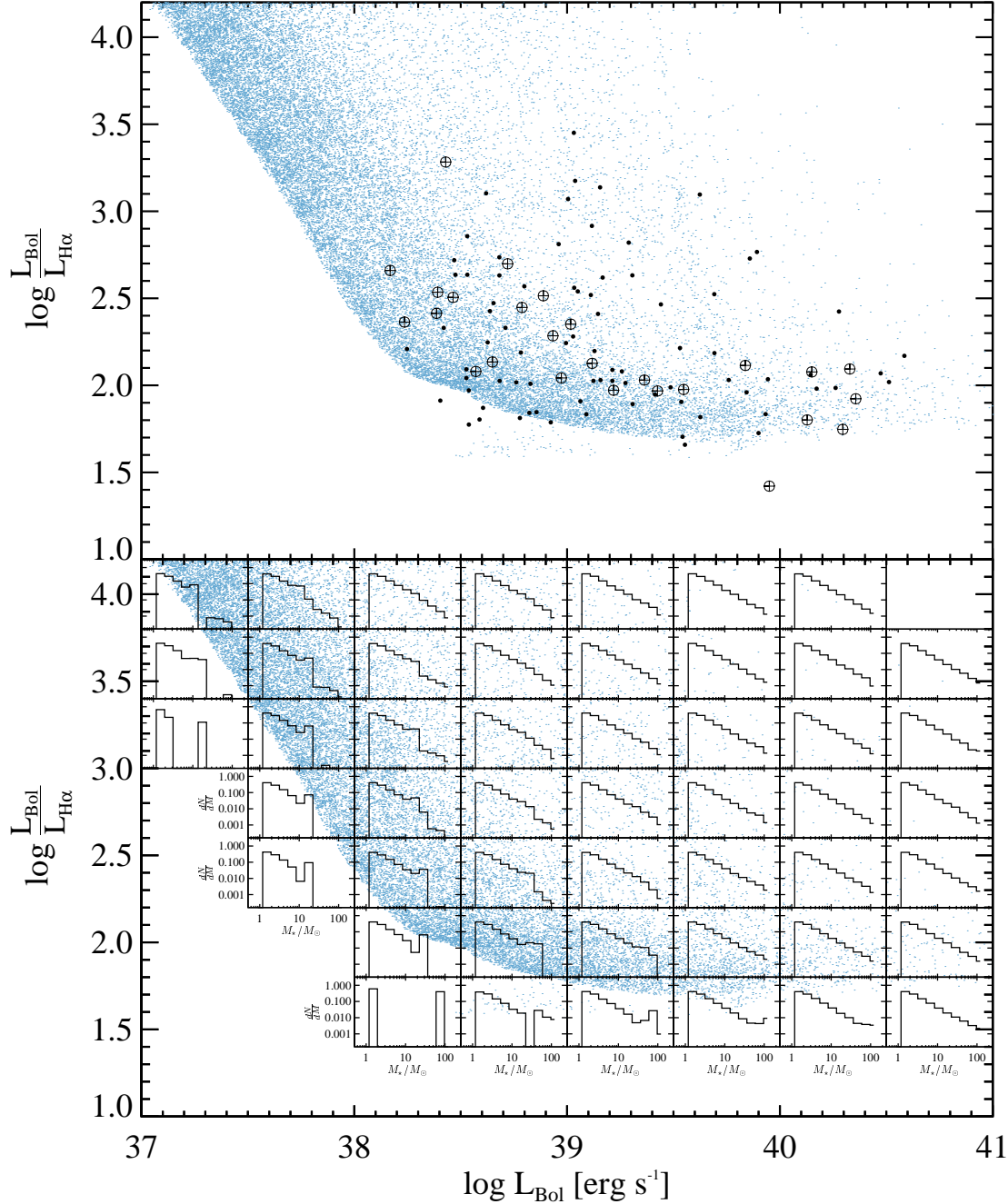}
\caption{ (\emph{top}) Here we present the birthline as first discussed by \cite{corbelli}.  Black points and
crosses are data from that paper with circle-crosses denoting their `clean' sample. Blue data points are clusters from SLUG. We
see that our models are in relatively good agreement with observations. (\emph{bottom}) We present overlays demonstrating the average IMFs in 
each region of the birthline plot.}
\label{fig:birthline}
\end{figure*}

\subsection{Cluster Birthline}
Another test of the photometry of clusters is to compare their H$\alpha$ luminosity to their bolometric luminosity.
Work by \cite{corbelli} has shown that newly born clusters lie along a birthline in this parameter space.
 In Fig.~\ref{fig:birthline} we compare the same models as Section \ref{sec:clusphot} (assuming $f_{esc}$ and $f_{dust}=0$) with their observational data and find good agreement. 
 Our theoretical
predictions differ slightly in the tilt of the locus of points from those
 by \cite{corbelli}, since we characterize the  properties of our stars in a different manner
  (making use
of stellar tracks rather than fitting formulae). 
To better demonstrate the origin of the birthline we also make use of SLUG's ability to keep track of the IMF of each individual cluster
(see bottom panel of Figure~\ref{fig:birthline}). Here we can see that the birthline from left to right forms a sequence of progressively 
more well-sampled upper ends of the IMF. Extremely rare deviants exist below the birthline where  more extremely massive ($>100M_\odot$)
stars are drawn than average, resulting in being born below the birthline.
Note that these rare clusters consisting of essentially isolated O stars have also been reported in the Milky Way \citep{dewit04,dewit05}
 and the SMC
\citep{oey04,lamb10} in numbers consistent with stochastic sampling of the IMF.

\subsection{Comparison with \textsc{SB99} }\label{sec:sb99}
A third obvious comparison for SLUG is \textsc{SB99} itself.
Since SB99 is widely used, it serves as a benchmark for SLUG.
Indeed, one of the motivations for making use of the SB99 tracks and SED matching algorithms
is that our code should be able to exactly reproduce SB99 if we select input parameters that place us in the continuously-sampled regime.
To that end we now present a variety of tests where we compare to SB99 to demonstrate that we can reproduce their results in the this
regime (the regime of a large galaxy-sized amount of stars).

\begin{figure}
\epsscale{1.}
\plotone{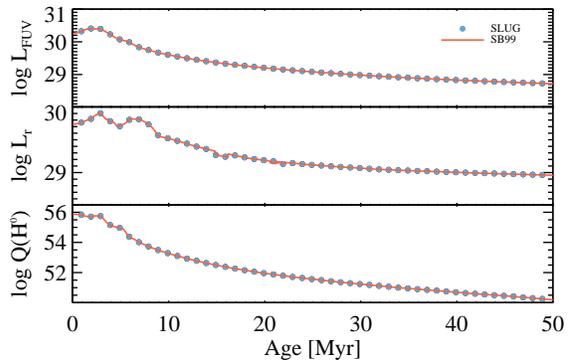}
\caption{A comparison of SLUG and \textsc{SB99} simulations of an instantaneous burst of  $10^9 M_\odot$.
 We find good agreement between the two in both the
absolute normalization of the fluxes as well as the time-dependent behavior. FUV and $r$ band fluxes are presented in units
of ergs/s/Hz while Q(H$^0$) is in units of photons/s. In the $L_r$ panel, one can see the effects of the discrete SED matching
techniques implemented by SB99 in the age ranges of 12-18 Myr and 20-22 Myr.}
\label{fig:sb99}
\end{figure}

\begin{figure}
\epsscale{1.}
\plotone{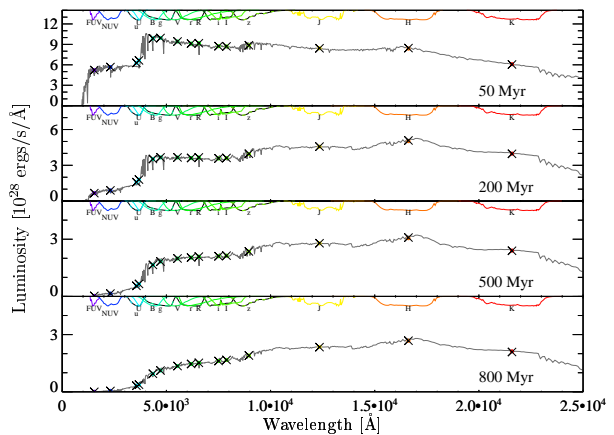}
\caption{A comparison of SLUG and SB99 photometry for an instantaneous burst of $10^9M_\odot$, evaluated at the ages indicated
in each panel.
The grey solid line represents the output spectrum from SB99 for such a population. The filled color circles show the
SB99 integrated fluxes for the filters available to SLUG. The black $\times$'s mark the SLUG photometry for the well-sampled model described
in section \ref{sec:sb99}.}
\label{fig:sb99sed}
\end{figure}

To compare the outputs of both SB99 and SLUG, we choose an instantaneous burst of star formation to demonstrate
the matching of the codes in both amplitude and time. 
We run a \textsc{SB99} model similar to our fiducial model (i.e. IMF slope of -2.35 from 1-120 $M_\odot$, solar metallicity, 
Padova+AGB tracks, Lej+SMI SEDs, and Maeder stellar wind models).  To meaningfully compare with SB99 we must
choose SLUG input parameters such that we are evaluating a population where SB99's approximations
are valid.  We therefore draw 
a very large instantaneous population of $10^9M_\odot$. To nullify any possible effects of our procedure
for populating the clusters, we ensure all clusters are very large by modifying the fiducial CMF to a restricted  range 
($10^6-2\times10^6 M_\odot$). We present the results in Fig.~\ref{fig:sb99}. It is evident that we are accurately able to reproduce \textsc{SB99}
in the well-sampled regime for integrated ``galaxy" properties. We match both the amplitude and time evolution
in all photometric bands.

This can also be seen by looking at the full SEDs. In Figure~\ref{fig:sb99sed},  we present photometry for
all 15 of the flux bands available for SLUG and compare with the spectra and integrated photometry produced by SB99 at a variety of time steps.
Again we are able to fully reproduce the photometric properties in the well-sampled regime from FUV to $K$-band.

In both these demonstrations, SLUG matches SB99 within 0.026 dex for all fluxes at all times.

\section{Stochasticity in Action}\label{sec:inaction}
Having demonstrated that SLUG can reproduce realistic clusters as well
as reproduce SB99's results, we now present outputs of SLUG in 
the stochastic regime.

\subsection{Effects on Coeval Populations }

\begin{figure}
\epsscale{0.8}
\plotone{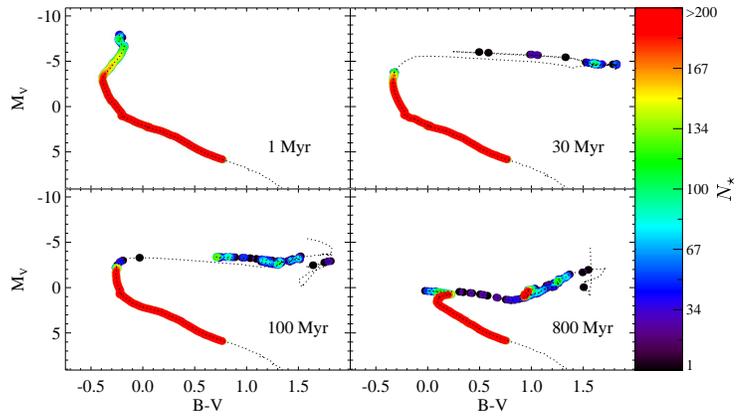}
\caption{CMDs of an instantaneous $10^5 M_\odot$ burst population at the ages indicated in each panel. 
Only stars more massive than 0.9 $M_\odot$ are binned in the CMDs.
The dotted lines show the corresponding theoretical isochrones. The SLUG CMD has been convolved with circular top-hat PSF solely
to improve visibility. The color bar denotes the number of stars in that region of the diagram.
 }
\label{fig:angst}
\end{figure}

Recent studies \citep[e.g.,][]{angst} have shown the wealth of information that 
can be obtained using resolved CMDs
of stars within a galaxy. For comparison with such studies in the stochastic
regime, SLUG produces binned 2 dimensional histograms that keep track
of the user's choice of color magnitude diagrams. Such diagrams allow us to 
directly characterize the effects of stochasticity in a coeval population.
In Figure~\ref{fig:angst}, we compare CMDs produced by SLUG for a $10^5 M_\odot$ instantaneous to

%
the theoretical isochrones from which they are produced. Aside from
demonstrating we accurately reproduce the tracks, we are able to see the effects of stochasticity
in leaving rapid phases of evolution unpopulated. Note that SLUG is capable of producing
such diagrams for any given SFH.

\subsection{Effects on Composite Populations }

\begin{figure*}

\begin{minipage}[b]{0.3\linewidth}
\centering
\epsscale{1.3}
\plotone{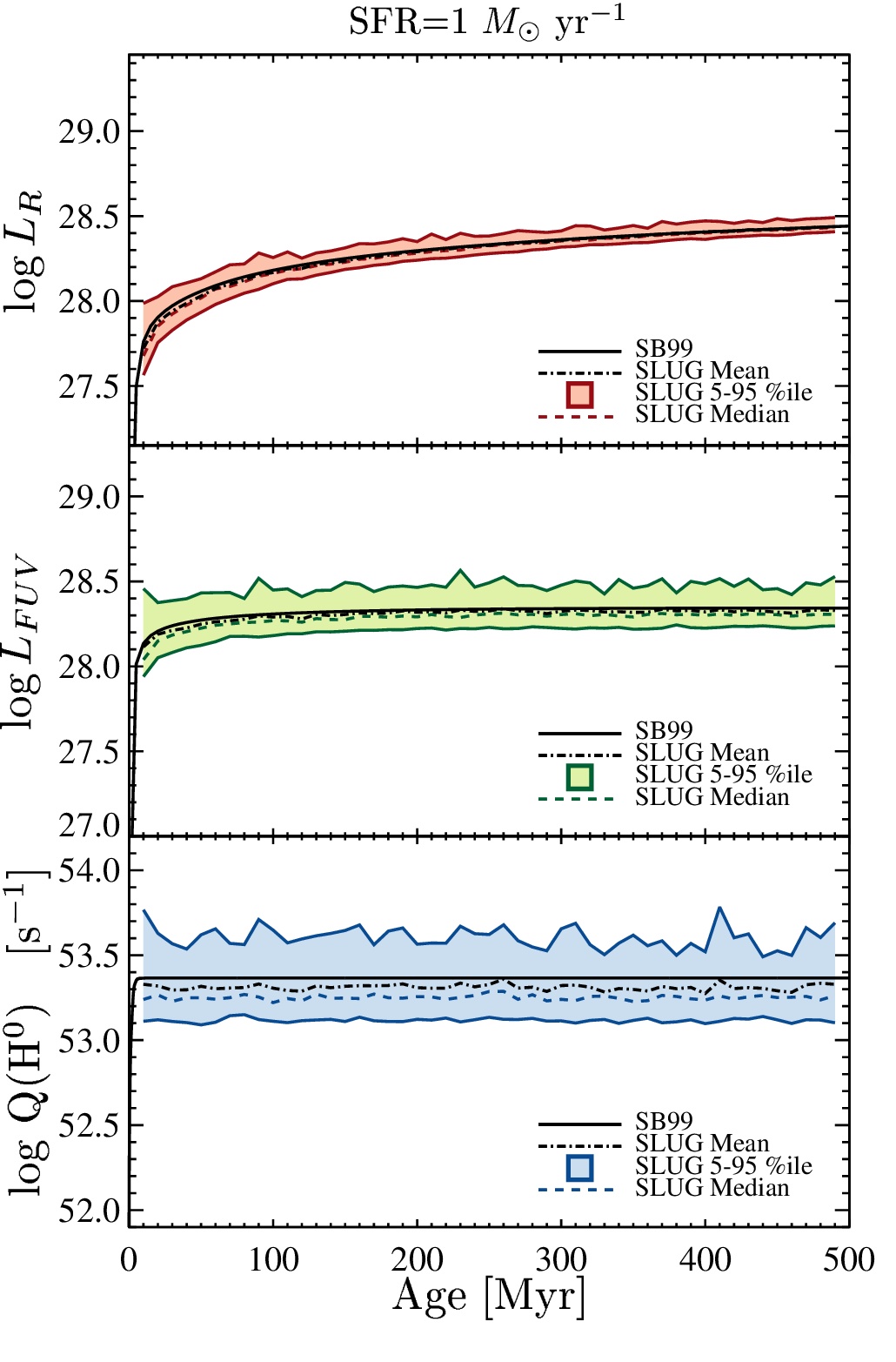}
\end{minipage}
\begin{minipage}[b]{0.3\linewidth}
\epsscale{1.3}

\plotone{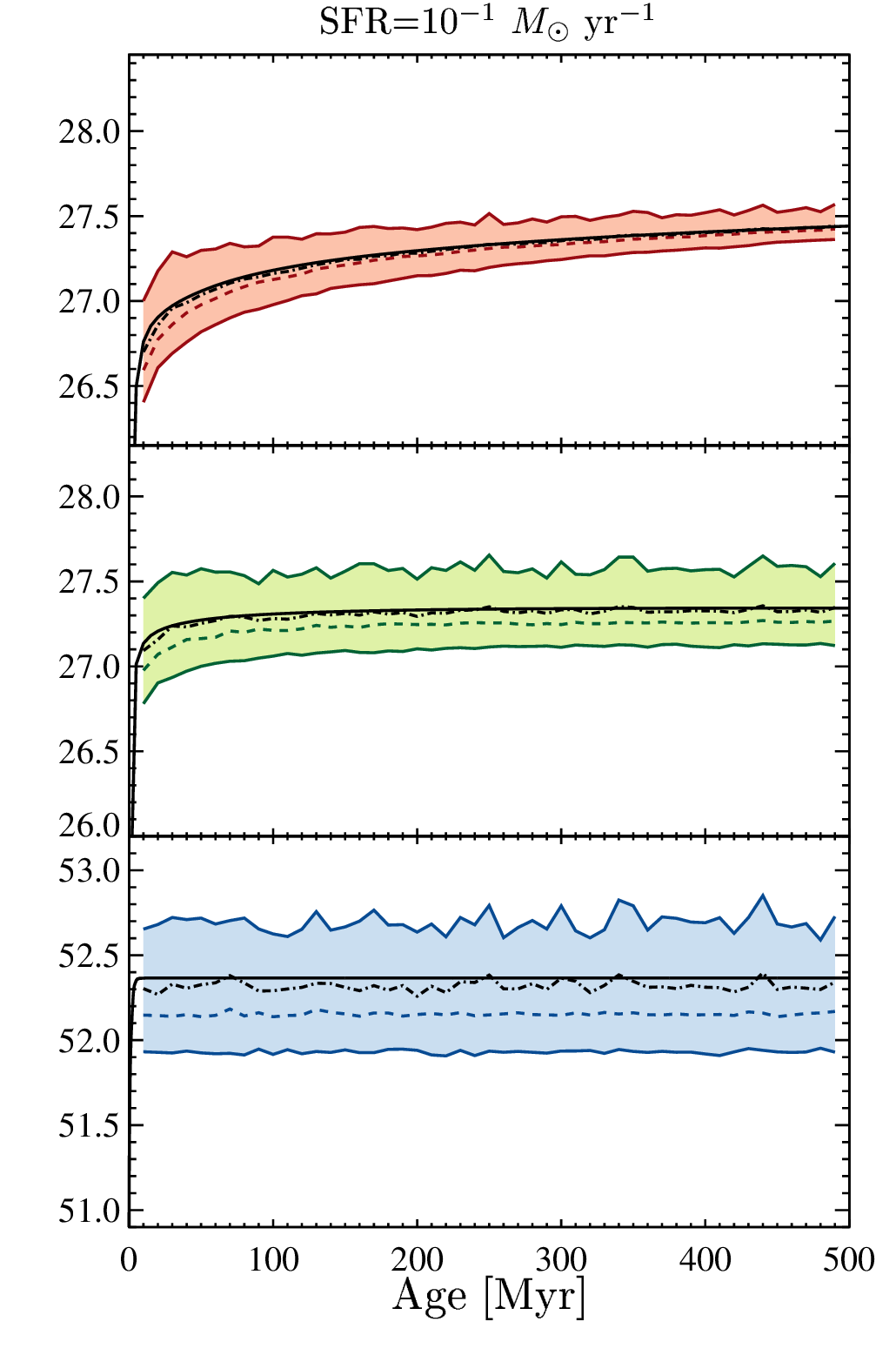}
\end{minipage}
\begin{minipage}[b]{0.3\linewidth}
\epsscale{1.3}

\plotone{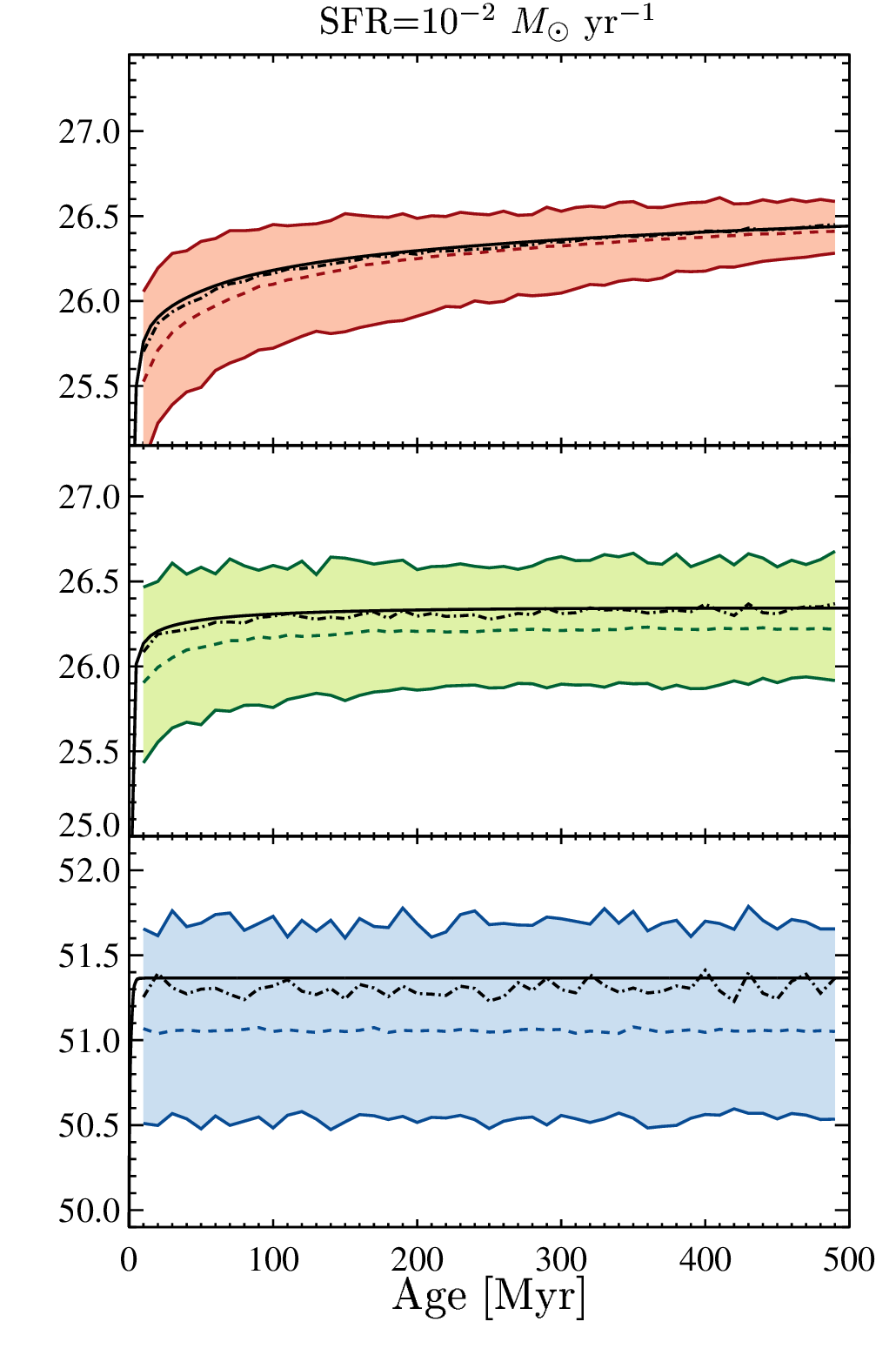}
\end{minipage}

\caption{
$R$-band, FUV, and ionizing photon luminosities vs. time for galaxies with constant SFRs of 
 1, $10^{-1}$, and $10^{-2}\ M_\odot$ yr$^{-1}$ as indicated. $R$-band and FUV
 luminosities are in units of erg s$^{-1}$ Hz$^{-1}$.
We compare
a fully sampled realization from \textsc{SB99} (solid black lines) 
with 100, 500, and 1000 realizations from SLUG for SFRs of 1, $10^{-1}$, and $10^{-2}\ M_\odot$ yr$^{-1}$ respectively.
The SLUG models are represented by their mean (black dash-dotted line), median (colored dashed line) and 5-95 percentile range 
(filled color region). Our SLUG models were set to only output every 10 million years. 
Note that the y-axis in each panel has been chosen to match the SFR, but always spans the same logarithmic interval. }

\label{fig:sb99stoch}
\end{figure*}

\begin{figure*}

\begin{minipage}[b]{0.3\linewidth}
\centering
\epsscale{1.3}
\plotone{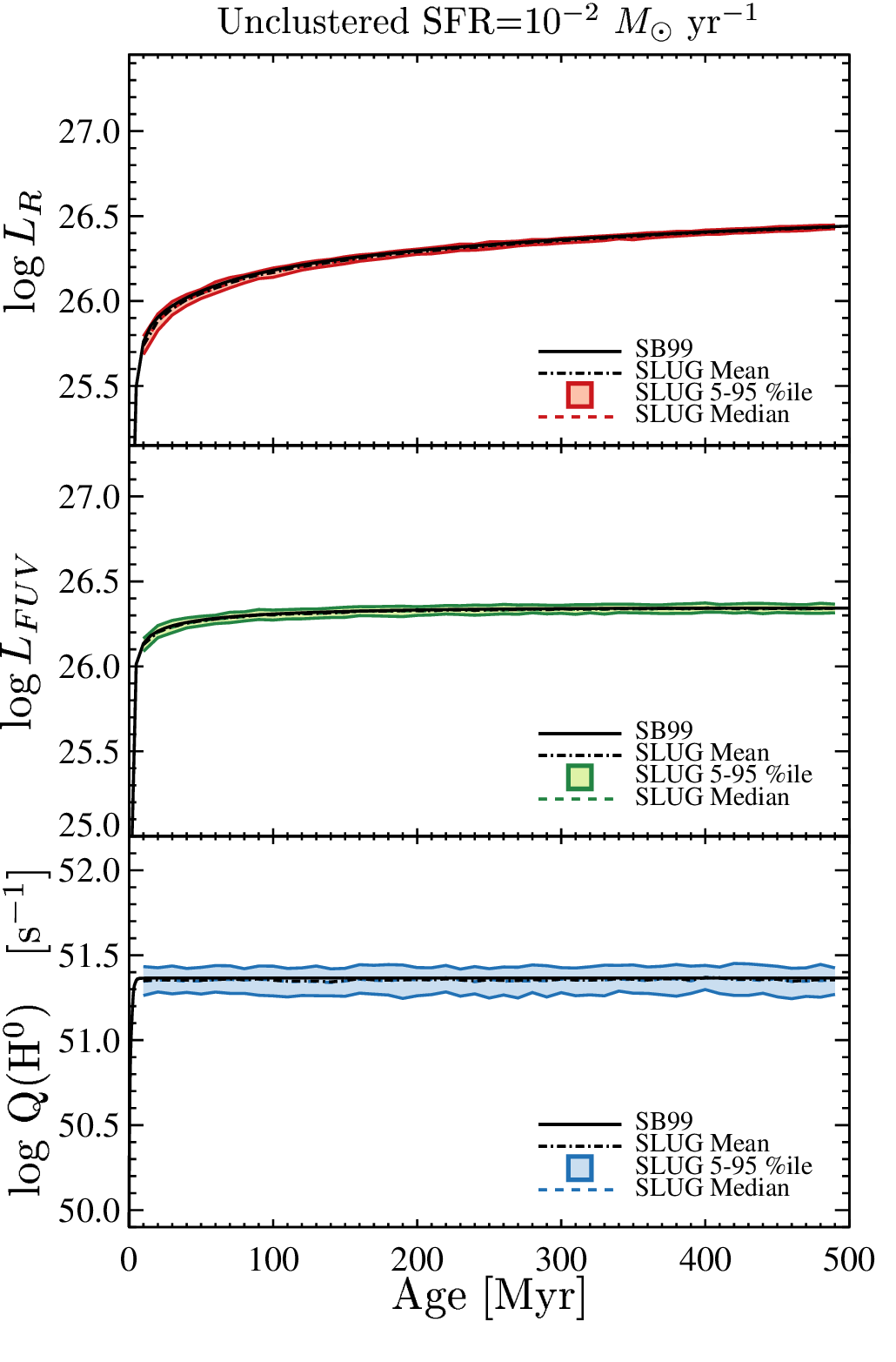}
\end{minipage}
\begin{minipage}[b]{0.3\linewidth}
\epsscale{1.3}
\plotone{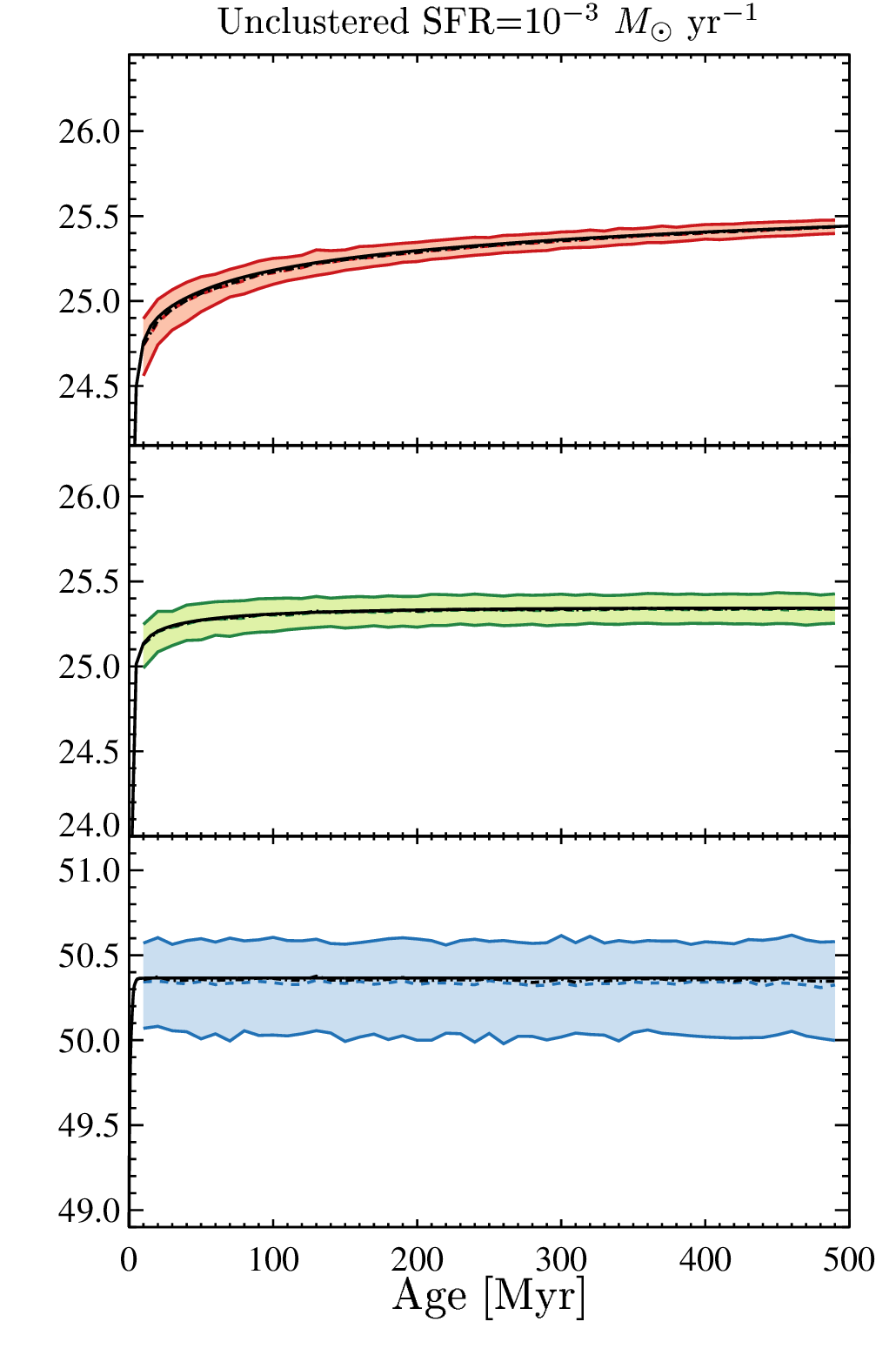}
\end{minipage}
\begin{minipage}[b]{0.3\linewidth}
\epsscale{1.3}
\plotone{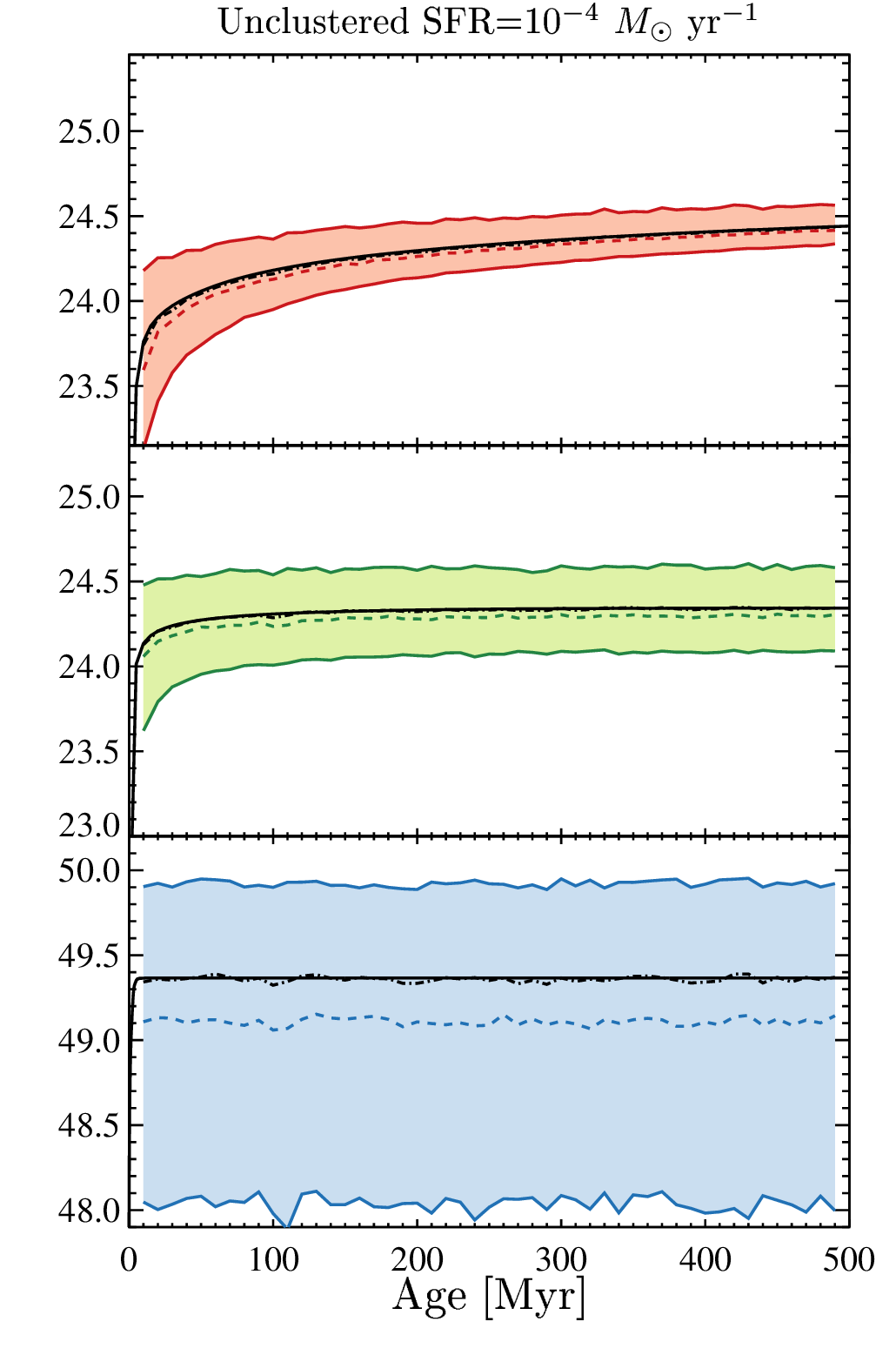}
\end{minipage}

\caption{Same as Figure \ref{fig:sb99stoch}, but this time made with unclustered star formation, and using lower SFR. Note the third panel
of Figure \ref{fig:sb99stoch} is the same SFR as the first panel of this figure. These figures were constructed 
with 100, 500,  and 1000 realizations
at SFRs of $10^{-2}$, $10^{-3}$, and $10^{-4}$ $M_\odot$ yr$^{-1}$ respectively.}

\label{fig:unclustered}
\end{figure*}

\begin{figure*}

\begin{minipage}[b]{0.3\linewidth}
\centering
\epsscale{1.3}
\plotone{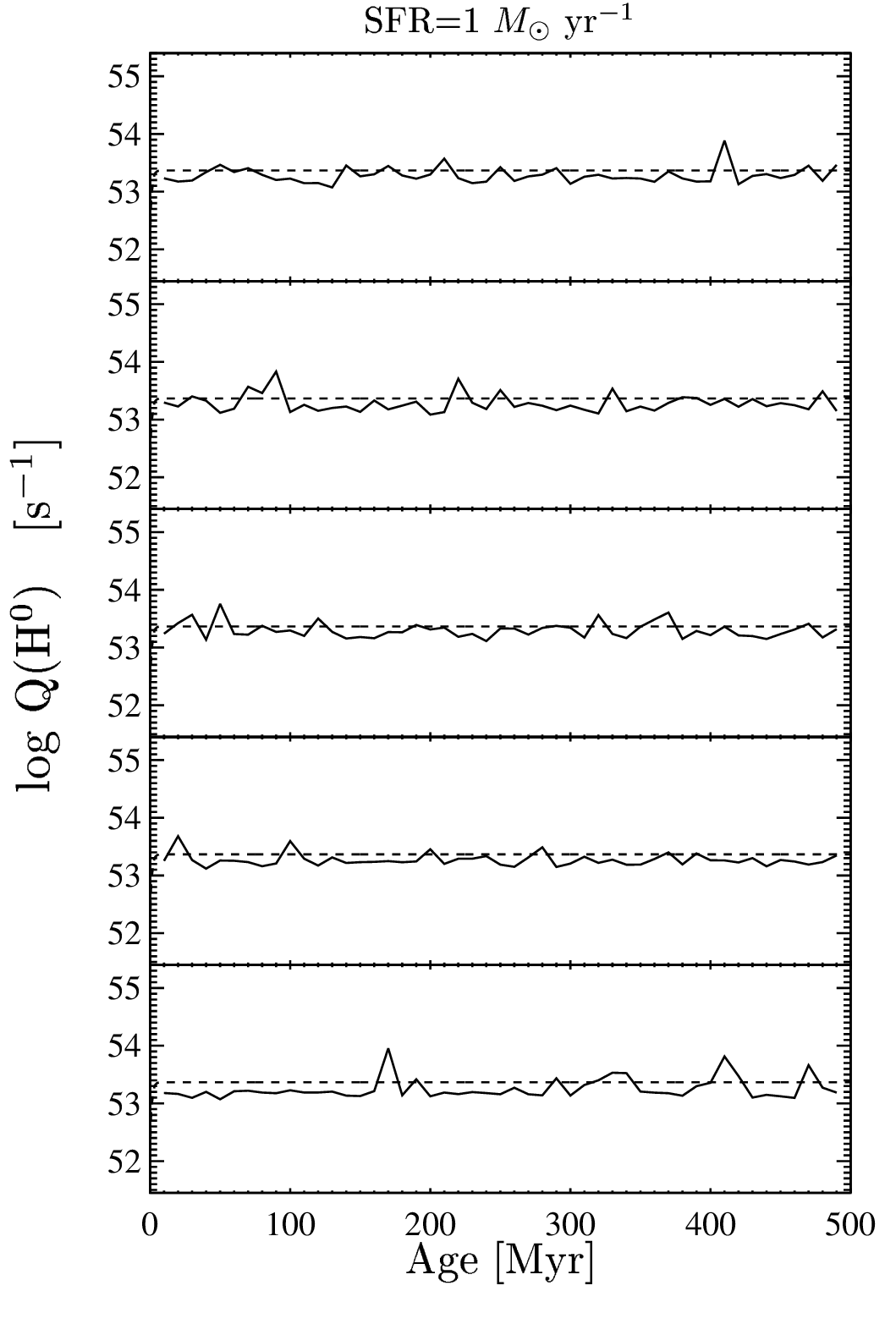}
\end{minipage}
\begin{minipage}[b]{0.3\linewidth}
\epsscale{1.3}
\plotone{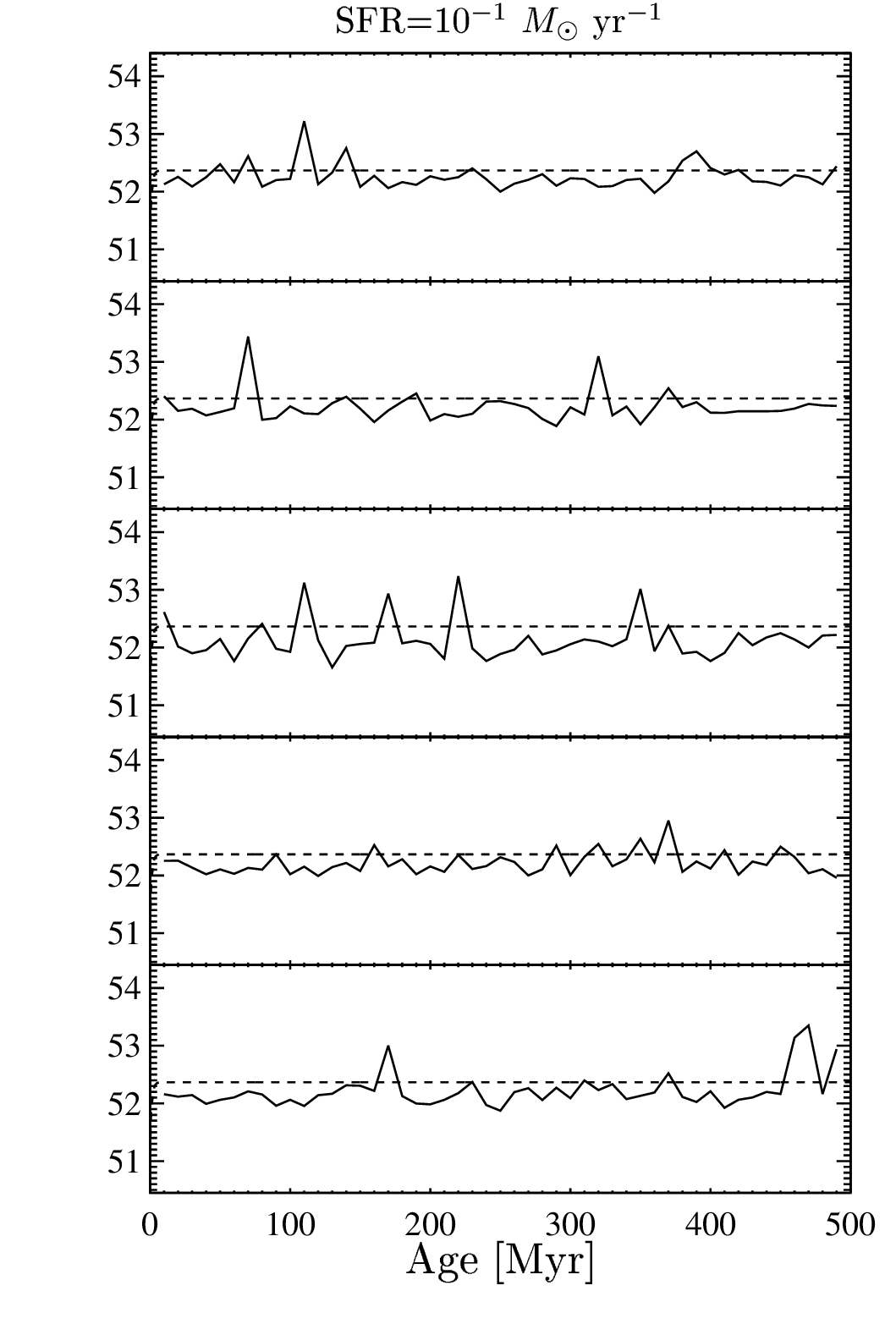}
\end{minipage}
\begin{minipage}[b]{0.3\linewidth}
\epsscale{1.3}
\plotone{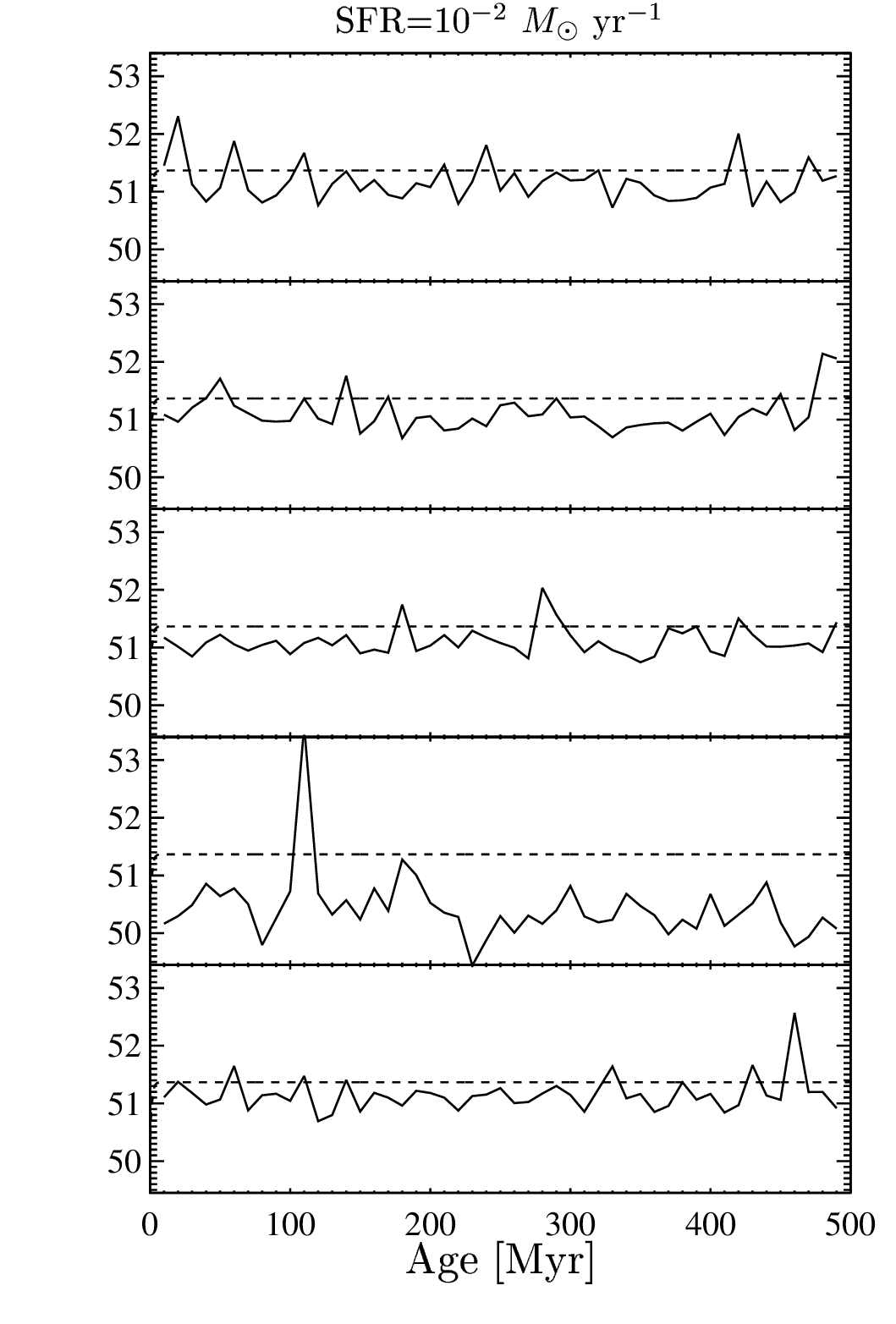}
\end{minipage}

\begin{minipage}[b]{0.3\linewidth}
\centering
\epsscale{1.3}
\plotone{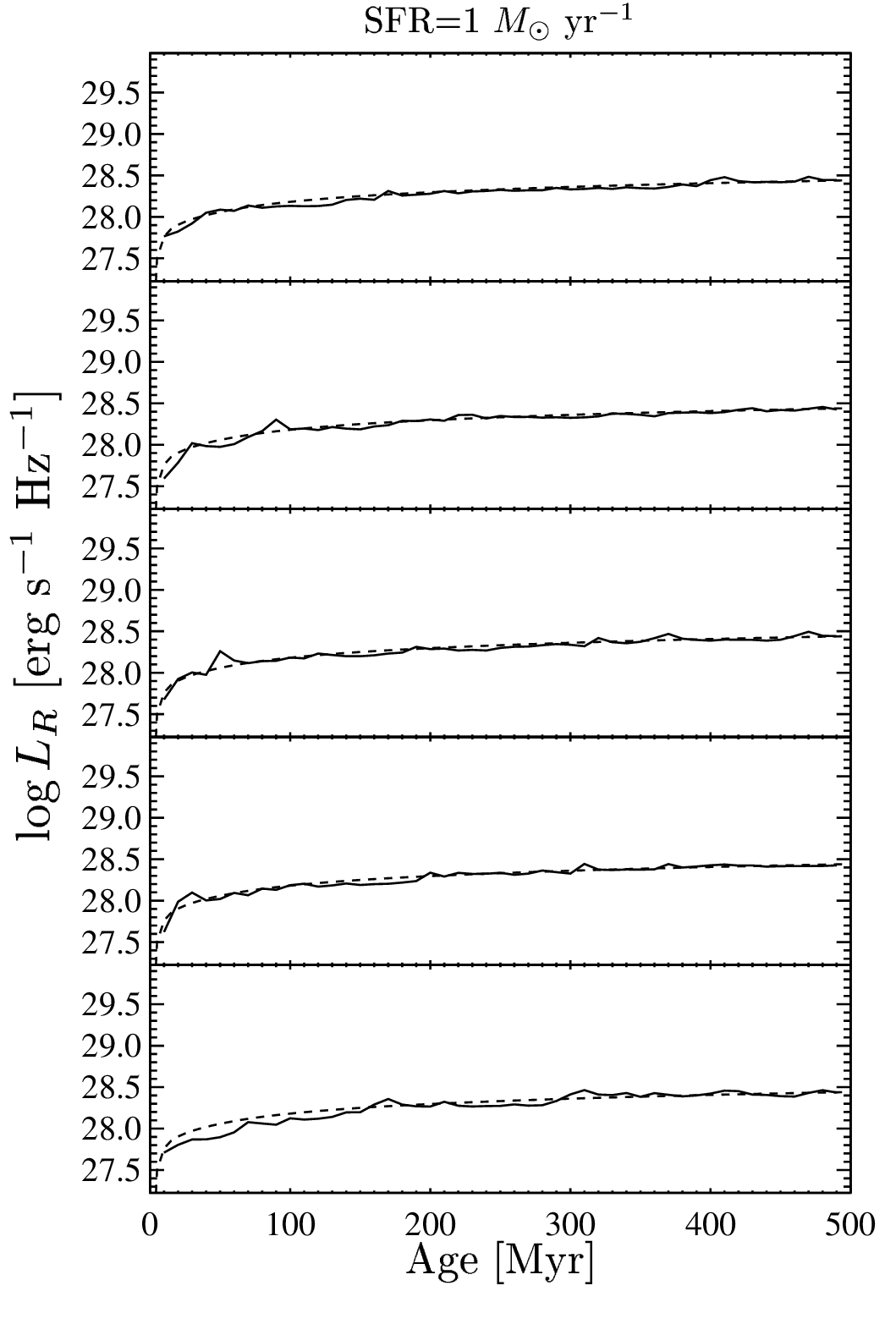}
\end{minipage}
\begin{minipage}[b]{0.3\linewidth}
\epsscale{1.3}
\plotone{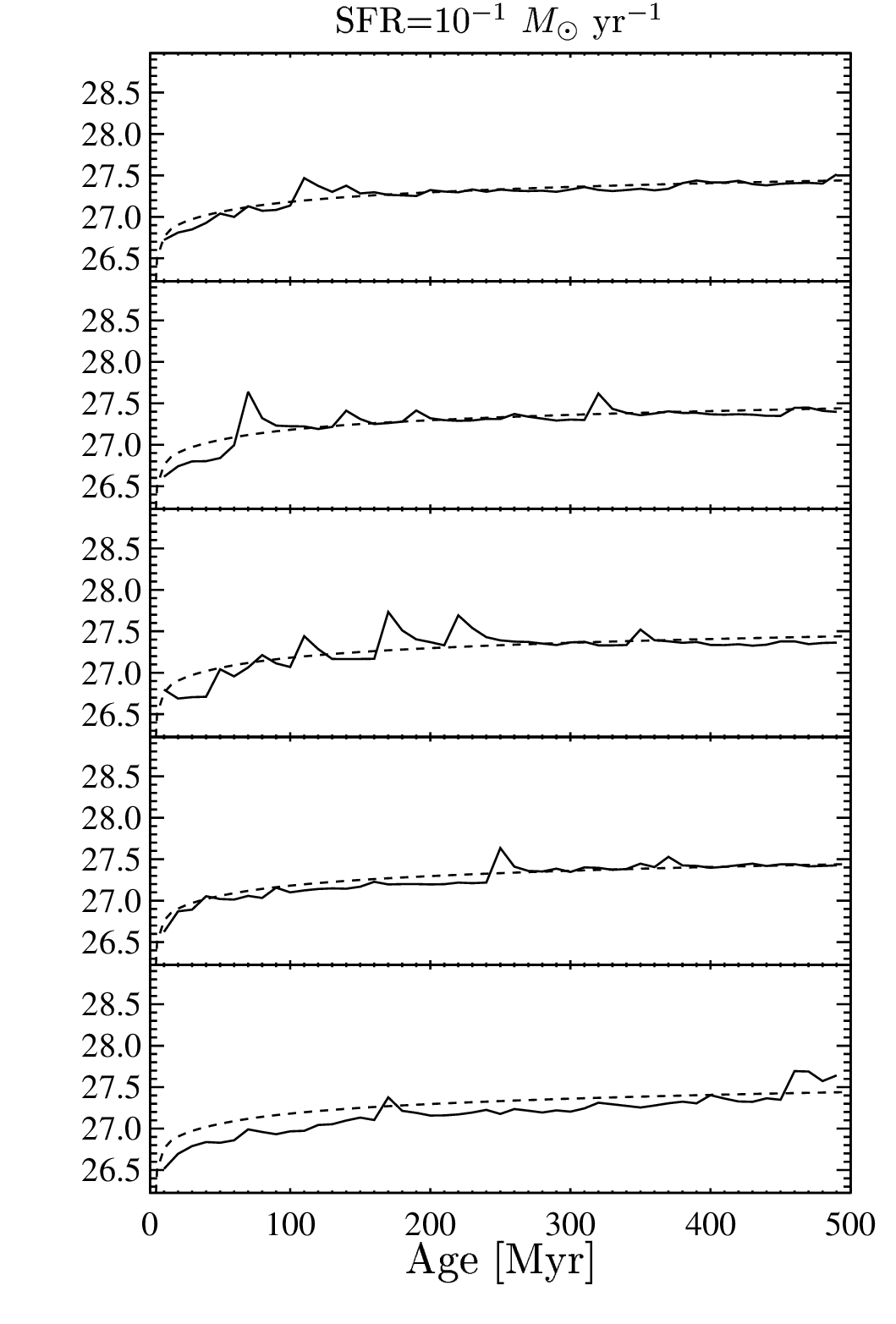}
\end{minipage}
\begin{minipage}[b]{0.3\linewidth}
\epsscale{1.3}
\plotone{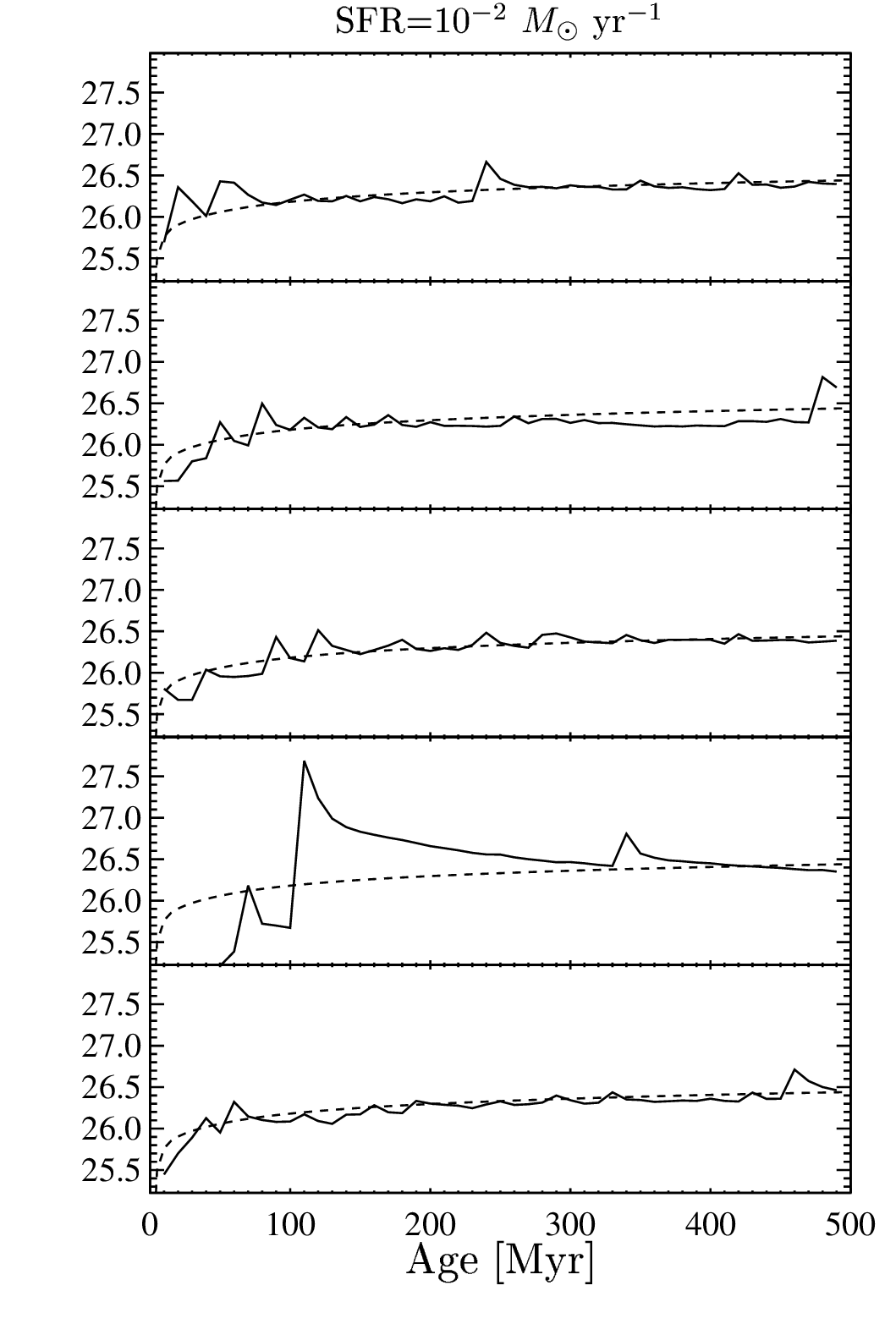}
\end{minipage}

\caption{Solid lines show the evolution of Q(H$^0$) and $R$ band luminosity 
for individual simulations with clustered star formation with 
SFRs of 1, $10^{-1}$, and $10^{-2}$ $M_\odot$ yr$^{-1}$.
Dashed lines show the SB99 prediction. 
Note that the y-axis in each panel has been chosen to match the SFR, but always spans the same logarithmic interval.}
\label{fig:tracks}
\end{figure*}

While individual clusters of stars can be treated as coeval, larger systems 
are intrinsically built of composite populations. One of the most basic
composite populations one can consider is a galaxy forming stars at 
a constant star formation rate. As discussed in Section \ref{sec:comp},
the value of the SFR will have a significant impact on the effects of stochasticity.

To demonstrate the differences that stochasticity makes, we compare SLUG realizations to those
of a well-sampled SB99 model. In Figure \ref{fig:sb99stoch}, we first
examine the luminosities for SFRs of 1, $10^{-1}$, and $10^{-2}$ $M_\odot$ yr$^{-1}$ with our fiducial values for the CMF
and cluster mass fraction.
For each SFR, we show the mean and median of the SLUG runs along with the 5 and 95 percentiles. 

One can clearly see an increase in fractional scatter as one decreases the SFR, which
can be attributed to the more bursty SFHs which are a result of the grouping of
age in massive clusters. This scatter appears at higher SFRs than predicted by our naive discussion in Section \ref{sec:comp} as
a direct result of the clustering. In fact, nearly all of the scatter seen in Figure \ref{fig:sb99stoch} is a result of the clustering
rather than sampling of individual clusters. This is most clearly demonstrated by Figure \ref{fig:unclustered}
which shows similar simulations but with completely unclustered star formation.
Without clustering the $10^{-2}$ $M_\odot$ yr$^{-1}$ models have approximately an order of 
magnitude less scatter 
in the 5-95 percentile range of the log of the luminosity. 
We see that the unclustered stochastic effects behave as predicted in Section \ref{sec:comp}
where the fractional scatter is small for SFRs $\sim 10^{-2}$ $M_\odot$ yr$^{-1}$ and quickly increases
as the SFR decreases \cite[also discussed in][]{mikiletter}.

For a demonstration of the effects of clustering, we present the tracks of a subset of 
individual stochastic realizations of clustered star formation in Figure \ref{fig:tracks}. One can see that the
Q(H$^0$) curves are less uniform than the $R$ luminosity. This is a direct result of the 
sensitivity of Q(H$^0$) to the youngest, most massive stars. One can also see that the scatter increases with decreasing
SFR as expected.
%
This is to be further discussed in da Silva et al. (in prep.) where we elaborate on the effects of stochastic star formation 
when one includes clusters.

\section{Summary}\label{sec:summary}
We introduce SLUG, a new code that correctly accounts for the effects of stochasticity 
(with caveats discussed in the text)
 by populating galaxies with stars and clusters of stars
and then following their evolution using stellar evolutionary tracks. Cluster disruption is
taken into account and a variety of outputs are created.


We present a series of tests comparing
SLUG  to observations and other theoretical predictions. SLUG is able to reproduce
the photometric properties of clusters from the \cite{larsen} catalog as well as the
\cite{corbelli} birthline. It can also reproduce the results of SB99 in the well-sampled regime.

Finally we present SLUG outputs in the stochastic regime and demonstrate the
flexibility of the code to address a variety of astrophysical problems with its variety of possible outputs.

SLUG is a publicly available code, and can be found at http://sites.google.com/site/runslug/.

\acknowledgements  
R.L.dS. is partially supported by 
an NSF CAREER grant (AST-0548180). The work of R.L.dS. is 
supported under a National Science Foundation Graduate 
Research Fellowship. MRK acknowledges support from: an 
Alfred P.~Sloan Fellowship; the National Science Foundation 
through grants AST-0807739 and CAREER-0955300; and NASA
 through Astrophysics Theory and Fundamental Physics grant 
 NNX09AK31G and a {\it Spitzer Space Telescope} theoretical research program grant.
 We would like to thank J. X. Prochaska for help in reading and providing input 
 on the early stages of this manuscript. We would like to thank
 F. Bigiel for encouraging us to create SLUG and useful conversations
 with J. Eldridge, C. Weidner, R. Bernstein, and J. Colucci. 

\clearpage
\appendix
\section{Implementation of IGIMF}\label{sec:igimf}

The IGIMF theory is a statement that the SFR controls the upper cutoff of the CMF and that
each cluster's mass changes the upper cutoff of the IMF in that cluster.

We implement the IGIMF following \cite{upigimf}. We use
the work of \cite{pflamm2008}, \cite{weid2005}, and \cite{weid2004b} 
to define the maximum cluster mass as
\begin{equation}
M_{ecl,max}=84793 \left( \frac{\left\langle  \mbox{SFR} \right\rangle }{M_\odot \mbox{ yr}^{-1}}\right)^{3/4},
\end{equation}
where $\left\langle  \mbox{SFR} \right\rangle$ is the time-average SFR.
Thus the SFR affects the upper cut off of the CMF. We determine the average
star formation rate over a time interval defined by the user (fiducially $10^7$ yr).

After a cluster mass has been drawn, we must adjust
the upper cutoff of the IMF that we use to draw
stars for that cluster. 
The relation between maximum stellar mass and  cluster mass ($m_{max}-M_{ecl}$)
has been studied  by \cite{weid2004} and \cite{ weid2010}. Following their treatment,
we solve a system of equations numerically for $m_{max}$ as a function of $M_{ecl}$.

\begin{figure*}
\epsscale{1.}
\plotone{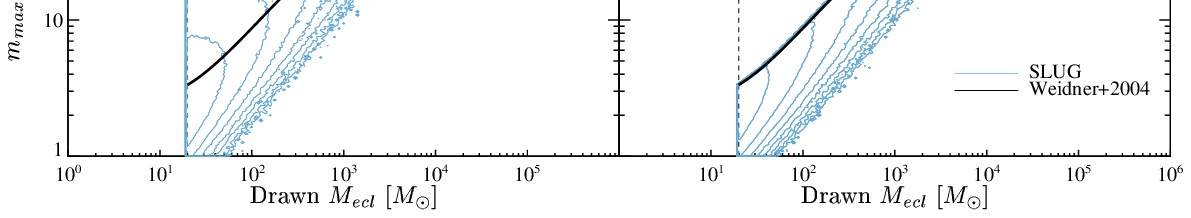}
\plotone{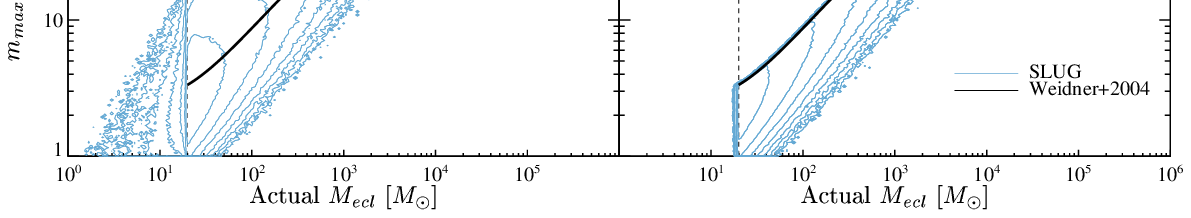}

\caption{
The mass of the largest star in a cluster vs. that cluster's mass for clusters created by SLUG for a \cite{Kroupa} IMF (\emph{left})
and the IGIMF (\emph{right}).
The black lines denote the analytic prediction of the maximum possible stellar mass in a cluster in the IGIMF model, the
 black dashed line notes the lower limit of the cluster mass function, and blue contours denote the location of SLUG models. 
Top panels show the maximum stellar mass as a function of the cluster mass drawn
from the CMF, while bottom panels show the same relation relative to the sum of the masses of all stars actually populating the clusters.
These two differ slightly-- see section \ref{sec:imf}.}
\label{fig:maxcl}
\end{figure*}

The first equation is simply a statement that the total cluster mass ($M_{ecl}$) is the integral of the 
distribution of masses ($\frac{dN}{dm}$) integrated from the lowest to highest mass star in the cluster:
\begin{equation}
M_{ecl}=\int_{m_{min}}^{m_{max}}m\frac{dN}{dm}dm.
\end{equation}
The next constraint is derived based on the statement that there is only one star in the cluster with mass equal to $m_{max}$.
Their choice
of implementation of this statement is as follows\footnote{ \cite{cervino} have pointed out
 that this expression does not equate to the 
logical statement mentioned above--it under-predicts the maximum mass in  63\% of cases. In fact, this formalism equates rather to the
statement that the expectation value of stars in the interval $m_{max}-m_{max,\star}$ is equal to 1. However this is the standard formalism of the
 IGIMF, so it is the formalism we implement.}:
\begin{equation}
1=\int_{m_{max}}^{m_{max,\star}}\frac{dN}{dm}dm
\end{equation}
where $m_{max,\star}$ is the maximum stellar mass possible.

In the specific case of a \cite{Kroupa} IMF, these equations reduce to the following \citep[taken from ][]{weid2004}.

\begin{equation}
1=k\left[ \left(\frac{m_H}{m_0} \right)^{\alpha_1} \left(\frac{m_0}{m_1} \right)^{\alpha_2} m_1^{\alpha_3} 
 \left(  \frac{m_{max,*}^{1-\alpha^3}}{1-\alpha_3}-\frac{m_{max}^{1-\alpha^3}}{1-\alpha_3}\right) \right]
\end{equation}

\begin{align}
\frac{M_{cl}}{k} &=  \frac{m_H^{\alpha_0}}{2-\alpha_0}    (m_H^{2-\alpha_0}-m_{low}^{2-\alpha_0})   +  
		\frac{m_H^{\alpha_1}}{2-\alpha_1}    (m_0^{2-\alpha_1}-m_{H}^{2-\alpha_1}) \nonumber \\
		&+           \frac{\left(\frac{m_H}{m_0}\right)^{\alpha_2} m_0^{\alpha_2}  }{2-\alpha_1}    (m_1^{2-\alpha_2}-m_{0}^{2-\alpha_2})+
		    \frac{\left(\frac{m_H}{m_0}\right)^{\alpha_2} \left(\frac{m_0}{m1}\right)^{\alpha_2}  m_1^{\alpha_3}}{2-\alpha_3}   
		    			 (m_{max}^{2-\alpha_3}-m_{0}^{2-\alpha_3})		
\end{align}
where
\begin{equation}
\begin{array}{ll}
\alpha_0=+0.30, & m_{low}=0.01 \\
\alpha_1=+1.30, & m_H=0.08\\
\alpha_2=+2.30, & m_0=1.00 \\
\alpha_3=+2.35, & m_{max,*}=120
\end{array}
\end{equation}
We fit a 6th order polynomial to the numerical solution to find:
\begin{equation}
\log_{10} m_{max}= \sum_{i=0}^6 a_i (\log_{10} M_{cl})^i
\end{equation}
where $a$=[      1.449,      -2.522,       2.055,     -0.616,     0.0897,   -0.00643,   0.000182].

We then use this upper mass limit to modify the standard \cite{Kroupa} IMF to fill in the stars for the cluster.
 Figure \ref{fig:maxcl} demonstrates the result. One can see that we are accurately applying the cutoff 
 to the IMF in the IGIMF.

\end{document}